\documentclass[man,noextraspace]{apa6}
\usepackage{IEEEtrantools}
\usepackage[english]{babel}
\usepackage[utf8x]{inputenc}
\usepackage{amsmath,amsthm,amssymb}
\usepackage{graphicx}
\usepackage{bm}
\usepackage[colorinlistoftodos]{todonotes}
\usepackage[authoryear]{natbib} 
\bibpunct{(}{)}{;}{a}{,}{,} 
\usepackage[flushleft]{threeparttable}
\usepackage{booktabs,caption}
\usepackage{colortbl}
\usepackage{graphicx}
\usepackage{pdflscape}
\usepackage{caption}
\usepackage{subcaption}
\usepackage{enumitem}
\usepackage{graphicx}
\usepackage{bm}
\usepackage{textcomp}
\usepackage{textgreek}
\usepackage{multicol}
\usepackage{setspace}
\usepackage{IEEEtrantools}
\usepackage{booktabs}
\usepackage{authblk}
\usepackage{outline}
\usepackage{mathrsfs}
\usepackage{mathdesign}
\usepackage{euscript}
\usepackage{amsbsy}
\usepackage{mathtools}
\usepackage{multirow}
\usepackage{rotating}
\usepackage[ruled,vlined]{algorithm2e}
\SetKwRepeat{Do}{do}{while}%
\usepackage{array}

\DeclareMathOperator*{\argmin}{argmin}
\DeclareMathAlphabet{\mathscrbf}{OMS}{mdugm}{b}{n}

\title{Penalized Estimation and Forecasting of Multiple Subject Intensive Longitudinal Data}
\shorttitle{MultiVAR}
\author[1]{Zachary F. Fisher}
\author[2]{Younghoon Kim}
\author[2]{Barbara L. Fredrickson}
\author[2]{Vladas Pipiras}

\affil[1]{The Pennsylvania State University}
\affil[2]{University of North Carolina at Chapel Hill}

\abstract{
Intensive Longitudinal Data (ILD) is an increasingly common data type in the social and behavioral sciences.  Despite the many benefits these data provide, little work has been dedicated to realizing the potential such data hold for forecasting dynamic processes at the individual level. To address this gap in the literature we present the \emph{multi-VAR framework}, a novel methodological approach allowing for penalized estimation of ILD collected from multiple individuals. Importantly, our approach estimates models for all individuals simultaneously and is capable of adaptively adjusting to the amount of heterogeneity present across individual dynamic processes. To accomplish this we propose a novel proximal gradient descent algorithm for solving the multi-VAR problem and prove the consistency of the recovered transition matrices. We evaluate the forecasting performance of our method in comparison with a number of benchmark methods and provide an illustrative example involving the day-to-day emotional experiences of 16 individuals over an 11-week period.  
}

\begin{document}
\makeatletter
\@ifundefined{@affil}{\def\@affil{~}{}}
\makeatother
\maketitle

\section{Introduction}
Intensive Longitudinal Data (ILD) is increasingly available to social and behavioral scientists. With this increased availability come new opportunities for modeling and predicting complex biological, behavioral and physiological phenomena.  Despite these new opportunities psychological researchers have not taken full advantage of promising opportunities inherent to this data, the potential to forecast psychological processes at the individual level.  To address this gap in the literature we present a novel modeling framework which addresses a number of topical challenges and open questions in the psychological literature on modeling dynamic processes. First, how can we model and forecast ILD when the length of individual time series and the number of variables collected are roughly equivalent, or when time series lengths are shorter than what is typically required for time series analyses? Second, how can we best take advantage of the cross-sectional (between-person) information inherent to most ILD scenarios while acknowledging individuals differ both quantitatively (e.g. in parameter magnitude) and qualitatively (e.g. in structural dynamics)?  Despite the acknowledged between-person heterogeneity in many psychological processes is it possible to leverage group-level information to support improved forecasting at the individual level? In the remainder of the manuscript we attempt to address these and other pressing questions relevant to the forecasting of multiple-subject ILD.

\subsection{Forecasting in Psychology}

Technological developments have significantly eased the burden of collecting intensive longitudinal data (ILD) for psychological researchers. This includes sensor-based physiological measurements, health and movement data, measures of behavioral and emotional states, as well as data from many other noisy and complex systems. Increased availability has brought with it the realization that ILD presents unique opportunities for psychological scientists looking to model, forecast and modify complex time-dependent processes. Despite this realization the lion's share of methods development within psychology has focused exclusively on explanation. That is, psychological researchers have primarily been concerned with the characterization of dynamic processes using a combination of theoretical knowledge and measures of model fit to guide model construction.

Despite this focus on explaining the past over predicting the future the development of modern forecasting methods specifically tailored to psychological data hold great promise for the field. For example, the accurate prediction of emotional and physiological states would be an invaluable tool for clinicians tasked with monitoring and intervening on individual behavior. Furthermore, accurate forecasts are helpful for identifying when and to whom an intervention should be applied. Forecasting also presents psychologists with a practical framework to assess conflicting evidence from empirical studies and competing causal theories.  In this paper we will focus specifically on forecasting daily measures of emotion dynamics and psychopathology, addressing some of the unique challenges inherent to this type of data. \\
\subsection{Vector Autoregressive Models in Psychology} 
In the social science and behavioral sciences Vector Autoregressive (VAR) models and their many flavors (e.g., Structural VAR, graphical VAR, time-varying VAR) have become a common approach for modeling ILD. VAR models have been used to model binge eating behaviors \citep{wild2010}, dynamics among mother-infant dyads \citep{ji2019, chen2020}, substance use patterns \citep{zheng2013}, and persistent depressive symptoms \citep{groen2019}, to name a few. VAR models are a natural fit for many idiographic analyses as they provide a concise interpretation of inter-variable relations. They are also visualized easily using path or network connection diagrams, and allow for the inclusion of many \emph{potentially relevant} variables. This is useful when theory does not give concrete guidance on whether a variable is related to the process under study.\\ 
VAR models are also a mainstay of forecasting in many fields. Consider econometrics, for example, where the widespread adoption of VAR models in the mid-1980s marked the beginning of a boon in forecasting practice \citep{allen2006}. These are just a few reasons why VAR models represent a natural jumping off point for applied social science researchers looking to apply forecasting methodologies in their work. However, there are a number of features common to ILD research which deserve additional attention in the context of VAR modeling.\\
The first issue we address was described by \citet{sims1980} as the "profligate parameterization" of the unrestricted or canonical VAR model. Indeed, the number of VAR parameters grows quadratically with each component series added to the system of equations. In this way the flexibility of the VAR model specification is also its Achilles' heel, there are a large number of unknown coefficients relative to the information available from the data. This imbalance can lead to overfitting the sample data and poor forecasting performance \citep{robertson2001}. This presents a potential problem for many ILD scenarios where time series lengths typically fall between $30$ and $100$ measurement occasions and many variables are collected (e.g., a $10-20$ item scale). In other words, employing the VAR model in applied research can be a delicate operation. One wants to include all relevant variables in a model to ensure the dynamics under study are well-captured, however, stringent theoretically-motivated restrictions are generally required to obtain a useful model.\\ 
The second issue our proposed method aims to address is that of multiple-subject ILD, and more specifically how to best utilize cross-sectional information when modeling intra-individual processes. This is a fundamental question in both psychology and time series analysis. In psychology, much attention has been paid to multilevel modeling as a means to synthesize time series data collected on multiple individuals \citep{bringmann2013a, epskamp2018a}. This approach is promising when the number of variables in the analysis is not large and individuals do not differ substantially in terms of their overall model structure. Another approach for leveraging cross-sectional information for multivariate time series modeling at the individual-level is Group Iterative Multiple Model Estimation \citep[GIMME;][]{gates2012}. The GIMME approach is built on the Structural-VAR (S-VAR) framework and is available in the \texttt{gimme} package \citep{gimme}. 
\subsection{Foundations of the Proposed Approach} 
With our approach we hope to retain the features of VAR modeling that are so attractive to social science researchers while confronting the problem of overparameterization.  To accomplish this we turn to methods that induce sparsity on the VAR parameter space through regularized estimation \citep{basu2015a,basu2015b}. Although it is also possible to address this issue by imposing some lower-dimensional structure on the data matrix, as in dynamic factor analysis \citep{molenaar1985,stock2002}, or by combining dimension reduction and VAR modeling \citep{bulteel2018b}, we focus our attention on the Least Absolute Shrinkage and Selection Operator \citep[LASSO; ][]{tibshirani1996} and adaptive LASSO \citep{zou2006} frameworks. Although originally developed for cross-sectional data these methods have been readily applied in the domain of multivariate time series analysis and a number of authors have found these methods to be successful in forecasting applications.  For example, \citet{li2014} found standard LASSO methods outperformed dynamic factor models in out-of-sample forecasting and \citet{medeiros2016a,kock2015} found the adaptive LASSO \citep{zou2006} to outperform standard forecasting approaches in both simulation studies and real-world data problems. \\
In the cross-sectional setting a number of authors have considered applying LASSO methods to data that arises from some fixed number of groups \citep{gross2016,ollier2017}.  These groups may represent different cohorts of individuals or different genres of movies, however, the underlying theme of these approaches is that we might learn more about each individual group or genre by structuring the combined data in some reasonable way. Here a sensible approach should return a pooled solution if in fact the underlying relations are identical across units of analysis, and return strictly unit-specific results if the units share little in common.  Most importantly, a sensible approach would be capable of operating in the gray area where some relations are common across units and others are unit-specific. \\
To the best of our knowledge the multi-VAR approach presented here is the first work that combines regularized estimation of time series models \citep{basu2015a,basu2015b} with the problem of supervised learning of multiple-group data  \citep{gross2016,ollier2017}. We believe this combination is exceptionally well suited many problems in the social sciences. In addition, we make a number of unique contributions to the existing literature. First, we prove a consistency result for our estimator in the proposed multiple-subject estimation problem.  Second, we propose a proximal gradient descent approach for solving the multiple-unit LASSO (standard and adaptive) problem based on the Fast Iterative Shrinkage-Thresholding Algorithm (FISTA) of \citet{beck2009}. Third, we evaluate the performance of our proposed method in a simulation study and a real data example from \citet{fredrickson2017} involving day-to-day emotional experiences. Finally, we provide a convenient R package for applied researchers looking to use the proposed methods \citep{multivar}.

\section{Estimating Vector Autoregressive Models}
We focus our attention on the multivariate time series, $\{\mathbf{X}_{t}\}_{t \in \mathbb{Z}} = \{(X_{j,t})_{j=1,\dots,d}\}_{t \in \mathbb{Z}}$. $\mathbf{X}_{t}$ is considered to follow a vector autoregressive model of order $p$, VAR($p$), if
\begin{equation}
\label{varp}
\mathbf{X}_{t} = 
  \boldsymbol{\Phi}_{1} \mathbf{X}_{t-1} + 
  \ldots + 
  \boldsymbol{\Phi}_{p} \mathbf{X}_{t-p} + 
  \boldsymbol{\varepsilon}_{t}, \quad t \in \mathbb{Z},
\end{equation}
for some $d \times d$ matrices $ \boldsymbol{\Phi}_{1}, \dots, \boldsymbol{\Phi}_{p}$ and a white noise series $\{\boldsymbol{\varepsilon}_{t}\}_{t \in \mathbb{Z}} \sim \text{WN}(\mathbf{0}, \boldsymbol{\Sigma}_{\boldsymbol{\varepsilon}})$ characterized by $\mathbb{E}(\boldsymbol{\varepsilon}_{t})=0$ and $\mathbb{E}(\boldsymbol{\varepsilon}_{t}\boldsymbol{\varepsilon}_{s}^{'})=0$ for $s \neq t$. For simplicity we assume $\mathbf{X}_{t}$ is of zero mean. Generally, a unique causal stationary solution to \eqref{varp} can be ensured by satisfying the stability condition given by $\text{det}( \boldsymbol{\Phi}(z)) \neq 0$, for $|z| \leq 1$, $z \in \mathbb{C}$, where $ \boldsymbol{\Phi}(z)=\mathbf{I}_{d} -\boldsymbol{\Phi}_{1}z - \ldots - \boldsymbol{\Phi}_{p} z^{p}$.

\subsection{Estimation of Unrestricted VAR Models}
It is common to estimate  \eqref{varp} using ordinary least-squares (OLS) regression,
\begin{equation}
(\widehat{\boldsymbol{\Phi}}_1,\ldots,\widehat{\boldsymbol{\Phi}}_p) = \argmin_{\boldsymbol{\Phi}_1,\ldots,\boldsymbol{\Phi}_p} \sum_{t=p+1}^T  \| \mathbf{X}_t - \boldsymbol{\Phi}_1 \mathbf{X}_{t-1} - \ldots - \boldsymbol{\Phi}_p \mathbf{X}_{t-p}\|_{2}^2,
\end{equation}
where $T$ is the sample size and $\| \cdot \|_{2}$ denotes the Frobenius (Euclidean) norm, which is equivalent to running component-wise regression on each of the $d$ VAR equations. In this case the estimate $\widehat{\boldsymbol{\Sigma}}_{\boldsymbol{\varepsilon}}$ is defined as the sample variance-covariance matrix of the residuals. When there are no restrictions on  $\boldsymbol{\Phi}$ the OLS estimates are asymptotically equivalent to those produced by Generalized Least Squares (GLS) \citep{zellner1962}. Under the assumption that  $\boldsymbol{\varepsilon}_{t} \sim \mathcal{N}(\mathbf{0}, \boldsymbol{\Sigma}_{\boldsymbol{\varepsilon}})$ are independent across $t$, the OLS estimates obtained by component-wise regression are also the Maximum Likelihood (ML) estimates \citep{lutkepohl2007}. These estimators are asymptotically normal under mild assumptions with explicit variance-covariance matrices.
 
A drawback of the unrestricted VAR model is the large number of parameters that must be estimated. In fact, the number of parameters scales quadratically as the number of component series increases. Assuming no mean structure, $pd^2 + d(d-1)/2$ model parameters need to be estimated for the unrestricted VAR($p$) model. This means that, for example, a VAR($1$) model with $10$ component series requires estimating $145$ parameters. With such a large parameter space it is likely that many of the estimated linear relationships in an unrestricted VAR model will be spurious and the regression matrix $\mathbf{X}^{'}\mathbf{X}$ ill-conditioned. Furthermore, when $(T-p)d < pd^2 + d(d-1)/2$ estimation via OLS is not possible.\\

\subsection{Estimation of Sparse VAR Models}

As a consequence of the dimensionality issues surrounding unrestricted VAR estimation, much attention has been paid to methods for reducing the VAR parameter space.  Ideally the selection of relevant series would be guided by theory. Unfortunately, the ease associated with many types of electronic data collection in the behavioral and social sciences has allowed for the collection of many repeated measures, all of which are hypothesized as relevant to the phenomena under study.  For this reason it is often difficult to prune variables \emph{a priori} when theory points to their inclusion. Several data-driven approaches have been presented in the literature to overcome this issue of high-dimensionality \citep{basu2015a,han2013}. In our current approach we assume sparsity of $\boldsymbol{\Phi}$ and use penalized estimation to recover the model parameters.  

\subsubsection{LASSO Estimation}

To set up the Least Absolute Shrinkage and Selection Operator \citep[LASSO; ][]{tibshirani1996} the VAR model and associated data are expressed in a regression form
\begin{equation} 
  \underbrace{
    \begin{bmatrix}
      \mathbf{X}_{p+1}^{'}\\ 
      \mathbf{X}_{p+2}^{'}\\ 
      \vdots\\ 
      \mathbf{X}_{T}^{'}\:\:
   \end{bmatrix}}_{\textstyle{ \mathcal{Y} }}
  =
  \underbrace{
    \begin{bmatrix}
      \mathbf{X}_{p}^{'}\quad & \ldots & \mathbf{X}_{1}^{'}\quad \\ 
      \mathbf{X}_{p+1}^{'} & \ldots & \mathbf{X}_{2}^{'}\quad \\ 
       \vdots & \ddots & \vdots \\ 
      \mathbf{X}_{T-1}^{'} & \ldots & \mathbf{X}_{T-p}^{'}\\ 
    \end{bmatrix}}_{\textstyle{ \mathcal{X} }}\:\:
  \underbrace{
    \begin{bmatrix}
      \boldsymbol{\Phi}_{1}^{'}\\ 
      \boldsymbol{\Phi}_{2}^{'}\\ 
      \vdots\\ 
      \boldsymbol{\Phi}_{p}^{'}\\ 
    \end{bmatrix}}_{\textstyle{ \mathcal{B}^{*} }}
  +
  \underbrace{
    \begin{bmatrix}
      \boldsymbol{\varepsilon}_{p+1}^{'}\\ 
      \boldsymbol{\varepsilon}_{p+2}^{'}\\ 
      \vdots\\ 
      \boldsymbol{\varepsilon}_{T}^{'}\\ 
    \end{bmatrix}}_{\textstyle{ \mathcal{E} }}
\end{equation}
or, equivalently, 
\begin{align} 
\mathrm{vec}(\mathcal{Y}) = &\ (\mathrm{I}_{d}\otimes\mathcal{X})\mathrm{vec}(\mathcal{B}^{*})+\mathrm{vec}(\mathcal{E}),\\
  \underbrace{\mathbf{Y}}_{Nd\times1} \label{eq5} 
  = &
  \underbrace{\mathbf{Z}}_{Nd\times q}\:
  \underbrace{\mathbf{B}^{*}}_{q\times1} +
  \underbrace{\mathbf{E}}_{Nd\times1},
\end{align}
where the star $*$ indicates the true parameter, $N=T-p$ and $q=pd^2$. Here, we assume that $\mathbf{B}^{*}$ is $s$-sparse (i.e. $\Sigma^{p}_{i=1}\|\mathrm{vec}(\boldsymbol{\Phi}_{i})\|_{0}=\|\mathbf{B}\|_{0}=\sum_{i=1}^{q}1_{\{B_{i}\neq0\}}=s$ where $\| \cdot \|_{0}$ is the $\ell_0$-norm). With this structure in place we can write the LASSO estimator as
\begin{equation}
\label{lasso}
\hat{\mathbf{B}} =
   \argmin_{\mathbf{B}\in \mathbb{R}^{q}} 
   \frac{1}{N} \|\mathbf{Y}-\mathbf{Z}\mathbf{B}\|^{2}_{2} + 
   \lambda\|\mathbf{B}\|_{1},
\end{equation}
where $\|\mathbf{B}\|_{1}=\Sigma^{q}_{i=1}|B_{i}|$ for $\mathbf{B}=(B_1,\dots,B_q)^{'}$ and $\lambda>0$ is the regularization penalty parameter. In \eqref{lasso} the scaling constant $\frac{1}{N}$ (corresponding to $\lambda$) sometimes takes the values $\frac{1}{2N}$, $1$ and $2$ depending on the convention.  Here $N=T-p$ refers to the time series length of a given individual in the sample. Changing the scaling context corresponds to a reparameterization of $\lambda$ and does not impact the estimation of \eqref{lasso}. Large values of $\lambda$ typically correspond to sparser solutions. 
\subsection{Multiple-Subject Penalized VAR}
Up to this point we have presented the VAR model and optimization problem in terms of a single multivariate time series. This was useful for describing the estimators, however, the majority of ILD and many psychophysiological applications involve observing the same variables across multiple subjects. With multivariate repeated measurements collected from multiple subjects we are now interested in estimating the sparse parameter vectors $\hat{\mathbf{B}}_{1},\dots,\hat{\mathbf{B}}_{K}$ for $K$ individuals.  Rarely if ever are the relationships among items strictly the same across any two individuals in the sample.  However, it is certainly possible and maybe even expected that certain qualitative aspects of a dynamic process are similar across individuals.  For this reason, strategies involving the estimation of $K$ separate LASSO problems are generally suboptimal. To overcome this limitation we propose the multi-VAR modeling framework for multivariate time series data collected from multiple subjects.\\
\subsection{The multi-VAR Approach}
The approach described herein relies on the following decomposition of $\mathbf{B}^{*}_{k}$,
\begin{equation}
\label{decomp}
\mathbf{B}^{*}_{k}= \boldsymbol{\mu}^{*}+ \boldsymbol{\Delta}^{*}_{k},\quad k=1,\ldots,K, 
\end{equation}
where $\boldsymbol{\mu}^{*} \in \mathbb{R}^{q}$ corresponds to the common effects across $K$ individuals and $\boldsymbol{\Delta}^{*}_{k} \in \mathbb{R}^{q}$ corresponds to the effects unique to individual $k$. Now, considering the regularization parameters $\lambda_{1}$ and $\lambda_{2, k}$, $k=1,\dots,K$, which govern the cross-sectional heterogeneity in our solution we can write the revised optimization problem as
\begin{equation}
\label{lasso1}
(\hat{\boldsymbol{\mu}}, \hat{\boldsymbol{\Delta}}_{1},\dots,\hat{\boldsymbol{\Delta}}_{K})=
   \argmin_{\boldsymbol{\mu}, \boldsymbol{\Delta}_{1},\dots,\boldsymbol{\Delta}_{K}} 
   \frac{1}{N} \sum^{K}_{k=1}\|\mathbf{Y}^{(k)}-\mathbf{Z}^{(k)}(\boldsymbol{\mu}+ \boldsymbol{\Delta}_{k})\|^{2}_{2} + 
   \lambda_{1}\|\boldsymbol{\mu}\|_{1} + \sum^{K}_{k=1} \lambda_{2,k}
   \|\boldsymbol{\boldsymbol{\Delta}_{k}}\|_{1} .
\end{equation}
As mentioned previously we prefer a sensible approach to handling multivariate time series data arising from multiple individuals, specifically in the case where the cross-sectional heterogeneity of the individual processes is unknown. If the individuals share very little in common, in terms of their time series, this approach should return essentially independent solutions. That is the results should be similar to what would be obtained from estimating $K$ separate VAR models. In \eqref{lasso}, larger values of the penalty parameter $\lambda$ will increasingly drive the corresponding coefficient matrix $\mathbf{B}$ to zero. Similarly, in \eqref{lasso1}, large enough values of $\lambda_{1}$ will shrink the common effect matrix, $\hat{\boldsymbol{\mu}}$ towards zero, and we will be left with the individual-specific effects $\hat{\mathbf{B}}_{k}=\hat{\boldsymbol{\Delta}}_{k}$. This would essentially produce results similar to those obtained from estimating $K$ independent VAR models. \\
Likewise, if the individual-level processes are essentially homogenous, a sensible approach would return results similar to estimating a single pooled model for all individuals in the sample. In this case we would expect that large enough values of $\lambda_{2,k}$ would drive the individual-specific transition matrices,  $\hat{\boldsymbol{\Delta}}_{k}$, towards zero, leaving only the common effect transition matrix to explain the individual-level results, $\hat{\mathbf{B}}_{k}=\hat{\boldsymbol{\mu}}$. Finally, if an individual's process has both common and unique components, a sensible approach would attempt to balance these contributions. In this case the penalty parameters, $\lambda_{1}$ and $\lambda_{2,k}$, are selected to optimally govern the contribution of both the common ($\hat{\boldsymbol{\mu}}$) and unique ($\hat{\boldsymbol{\Delta}}_{k}$) effects to each individual's dynamics ($\hat{\mathbf{B}}_{k}$). \\
Using the decomposition presented in $\eqref{decomp}$ it is also possible to rewrite the right-hand side (RHS) of \eqref{lasso1} such that $\boldsymbol{\mu}$ only appears in the penalty term as
\begin{align}
\label{lasso2}
   & \argmin_{\boldsymbol{\mu}, \mathbf{B}_{1},\dots,\mathbf{B}_{K}} 
   \frac{1}{N} \sum_{k=1}^{K}\|\mathbf{Y}^{(k)}-\mathbf{Z}^{(k)}\mathbf{B}_{k}\|_{2}^{2}+
    \lambda_{1}\|\boldsymbol{\mu}\|_{1}+\sum_{k=1}^{K}\lambda_{2,k}\|\mathbf{B}_{k}-\boldsymbol{\mu}\|_{1}   \nonumber \\
  =& \argmin_{\boldsymbol{\mu}, \mathbf{B}_{1},\dots,\mathbf{B}_{K}} 
   \frac{1}{N} \sum^{K}_{k=1}\|\mathbf{Y}^{(k)}-\mathbf{Z}^{(k)}\mathbf{B}_{k}\|^{2}_{2} + 
   \lambda_{1}\left( \|\boldsymbol{\mu}\|_{1} + \sum^{K}_{k=1}  \frac{\lambda_{2,k}}{\lambda_{1}}
   \|\mathbf{B}_{k}-\boldsymbol{\boldsymbol{\mu}}\|_{1} \right).
\end{align}
To simplify the following discussion let $r_{k}=\lambda_{2,k} / \lambda_{1}$ for $k=1,\dots,K$. Now, it is important to note, as in \citet{gross2016}, that any choice of the regularization parameters, $\frac{\lambda_{2,1}}{\lambda_{1}},\dots,\frac{\lambda_{2,K}}{\lambda_{1}}$, and coefficients $B_{1,j},\ldots,B_{K,j}$ identifies a specific solution for the common effects  in $\boldsymbol{\mu}$ where $\mu_{j}$ is the weighted and shrunken median of $B_{1,j},\dots,B_{K,j}$ as in \citet{ollier2017} . \\
Indeed, the penalty term in \eqref{lasso2} is separable in its $q$ parameters such that we can consider a single explanatory coefficient $B_{k,j}$ and associated weight $r_{k}$ for $k=0,\dots,K$. Using this specification we can rewrite the penalty term in \eqref{lasso2} as the sum of the generic one-dimensional unconstrained optimization problem
\begin{equation}
\label{lasso3}
   \argmin_{\mu_{j}}\sum_{k=0}^{K}r_{k}|B_{k,j}-\mu_{j}|.
\end{equation}
Implicitly we set $r_{0} = 1$ and  $B_{0,j}=0$ to match the penalty construction in \eqref{lasso2}. Expressed as in \eqref{lasso3}, the solution $\hat{\mu}_{j}$ becomes a properly weighted median of $(B_{0,j}=0,B_{1,j},\dots,B_{K,j})$. In this setting a number of scenarios relevant to applied researchers are worth considering, as discussed in \citet{gross2016}. First, if $r_k = r$, $k = 1,\ldots, K$ and $r \in (\frac{1}{K}, \frac{1}{K-2})$ the group effect $\hat{\mu}_{j}$ will be nonzero if and only if all $B_{k,j}$ are of the same sign, in which case it will be equal to the minimum value of $(B_{1,j},\dots,B_{K,j})$. This means for the group effect to exist it must be present for all individuals in the sample and only then will deviations from the group be captured in the individual $(B_{1,j},\dots,B_{K,j})$. Second, if $\Sigma_{k=1}^{K} r_{k} < 1$ we are guaranteed the group effect $\hat{\mu}_{j}$ will equal zero and the problem will resemble fitting $K$ individual penalized VAR models.  Third, if $r_k = r$, $k = 1,\ldots, K$ and $r > 1$ the group effect $\hat{\mu}_{j}$ will effectively be the median of $(B_{1,j},\dots,B_{K,j})$.  In general, the weights $r_{k}$ in the above minimization problem can be understood as a penalty applied to idiosyncratic dynamics (coefficients) not shared by the entire sample. 
\subsection{The Adaptive multi-VAR Approach}
It is also possible to develop an adaptive-LASSO \citep{zou2006}  version of the multi-VAR approach for the VAR model by minimizing the objective function  
\begin{equation}
\label{adalasso}
   \frac{1}{N} \sum^{K}_{k=1}\|\mathbf{Y}^{(k)}-\mathbf{Z}^{(k)}(\boldsymbol{\mu} + \boldsymbol{\Delta}^{(k)})\|^{2}_{2} + 
   \lambda_{1}\left(\boldsymbol{\omega} \|\boldsymbol{\mu}\|_{1} + \sum^{K}_{k=1}  \frac{\lambda_{2,k}}{\lambda_{1}}\boldsymbol{\nu} _{k}
   \|\boldsymbol{\Delta}^{(k)}\|_{1} \right),
\end{equation}
where $\omega_{j}=1/|\tilde{B}_{\ell_{j},j}|$ and $\nu_{k,j}=1/|\tilde{B}_{k,j}-\tilde{B}_{\ell_{j},j}|$ with $\tilde{B}_{k,j}$ and $\tilde{B}_{\ell_{j},j}$ defined next. For each of the $k$ individuals in the sample the estimate $\tilde{\mathbf{B}}_{k}=(\tilde{B}_{k,j})$ of $\mathbf{B}_{k}$ can be obtained using maximum likelihood or OLS when the number of time points for each individual ($Nd$) exceeds the number of variables  ($pd^{2}+d(d-1)/2$), or from \eqref{lasso2} when this condition is not met.  In addition, $\tilde{B}_{\ell_{j},j}$ can be taken as the median coefficient estimate for variable $j$ across all $K$ individuals such that $\tilde{B}_{\ell_{j},j}=\text{median}(\tilde{B}_{1,j},\dots,\tilde{B}_{K,j})$. A benefit of the adaptive LASSO approach in comparison to \eqref{lasso1} is that we are able to weight the $\ell_{1}$ penalty. By weighting the $\ell_{1}$ penalty we are able to help ensure coefficients we might expect to be prominent in the model (based on a consistent first stage estimator, such as OLS) receive smaller penalties. In certain contexts this helps to reduce the bias of the LASSO estimator and can provide a number of benefits, including consistency in both variable selection and  parameter estimation \citep{zou2006}.
A nice property of this approach is that we can reexpress \eqref{lasso2} and \eqref{adalasso} as a weighted LASSO problem, namely,

\begin{equation}
\label{adalasso2} 
  \underbrace{
    \begin{bmatrix}
      \mathbf{Y}^{(1)}\\ 
      \mathbf{Y}^{(2)}\\ 
      \vdots\\ 
      \mathbf{Y}^{(K)}\:\:
   \end{bmatrix}}_{\textstyle{ \mathscrbf{Y} }}
  =
  \underbrace{
    \begin{bmatrix}
      \mathbf{Z}^{(1)} & \mathbf{Z}^{(1)} & \mathbf{0} & \ldots & \mathbf{0} \\ 
      \mathbf{Z}^{(2)} & \mathbf{0} & \mathbf{Z}^{(2)} & \ldots & \mathbf{0} \\ 
      \vdots & \vdots & \vdots & \ddots & \vdots \\ 
      \mathbf{Z}^{(K)} & \mathbf{0} & \mathbf{0} &\dots & \mathbf{Z}^{(K)} \\ 
    \end{bmatrix}}_{\textstyle{ \mathscrbf{Z} }}
  \underbrace{
    \begin{bmatrix}
      \boldsymbol{\mu^{*}}\\ 
      \boldsymbol{\Delta}^{(1)*}\\ 
      \vdots\\ 
      \boldsymbol{\Delta}^{(K)*}\\ 
    \end{bmatrix}}_{\textstyle{ \boldsymbol{\theta} }^{*}}
  +
  \underbrace{
    \begin{bmatrix}
      \mathbf{E}^{(1)}\\ 
      \mathbf{E}^{(2)}\\ 
      \vdots\\ 
      \mathbf{E}^{(K)}\:\:
   \end{bmatrix}}_{\textstyle{ \mathscrbf{E} }},
\end{equation}
where the criterion we are now concerned with minimizing is given by
\begin{equation}
\label{lassobig}
   \argmin_{\boldsymbol{\theta}} 
   \frac{1}{N} \|\mathscrbf{Y}-\mathscrbf{Z}\boldsymbol{\theta}\|^{2}_{2} + 
   \lambda_{1}\|\boldsymbol{\theta}\|_{1,\mathbf{w}} 
\end{equation}
and $\|\boldsymbol{\theta}\|_{1,\mathbf{w}} = \sum_{i}w_{i}|\theta_{i}|$ with $\mathbf{w}^{'}=(\mathbf{1}^{'}_{d}, (\lambda_{2,1}/\lambda_{1})\mathbf{1}^{'}_{d},$$\dots,(\lambda_{2,K}/\lambda_{1})\mathbf{1}^{'}_{d})$ for \eqref{lasso2} and $\mathbf{w}^{'}=(\boldsymbol{\omega}^{'},(\lambda_{2,1}/\lambda_{1})\boldsymbol{\nu}_{1}^{'},$$\dots,(\lambda_{2,K}/\lambda_{1})\boldsymbol{\nu}_{K}^{'})$ for \eqref{adalasso}. \\

It is worth nothing that the design matrix $\mathscrbf{Z}$ in \eqref{adalasso2} is not of full column rank even if the number of observations per subject will exceed the number of parameters. In particular, OLS for \eqref{adalasso2} is not feasible. Yet, under sparsistency, results on consistency and sparsistency for LASSO estimation are available as discussed in the appendix below.

\section{Computational Algorithm and Estimation}
Solving \eqref{lassobig} requires iterative methods as the $\ell_{1}$ penalty is not differentiable and no analytic solutions exist.  A popular schema for solving penalized regression problems  is coordinate descent as popularized by \citet{friedman2010}.  Coordinate descent has proved to be an exceedingly effective algorithm for exploiting the sparsity of the coefficient vector structure, partly because it moves parameters one at a time. Coordinate descent is easier to implement than many competing approaches and this has likely also contributed to its popularity.  Another class of methods for solving \eqref{lassobig} fall under the umbrella of proximal gradient descent. Unlike coordinate descent, proximal gradient descent moves all the parameters of a model at once, and may provide efficiency gains for certain types of problems, such as the estimation of high-dimensional VAR models \citep{nicholson2017}. We have chosen to implement our approach in the proximal gradient framework due to these desirable qualities, as well as the generality of the proximal framework to a wide-range of time-series optimization problems applicable to the multi-VAR framework. In the remainder of this section we introduce the proximal gradient descent algorithm we have implemented for solving \eqref{lassobig} and describe a number of useful modifications for enhancing computational efficiency.\\

Proximal algorithms have proved incredibly useful in the fields of statistics, machine learning and image processing for solving complex optimization problems involving composite objective functions, such as the one presented in \eqref{lasso2}. In fact, many methods commonly used in psychometric research, such as Expectation Maximization (EM), Majorization-Minimization (MM) and Iteratively Reweighted Least Squares (IRLS), can be shown to be proximal algorithms \citep{polson2015}. Broadly, a proximal algorithm refers to any algorithm where a proximal operator is applied to a subproblem of a larger optimization routine, often in a nonsmooth setting where the aim is simplifying the problem of interest.  It is beyond the scope of the current work to describe the proximal operator itself in any generality, however, a detailed treatment of proximal operators and algorithms is given by \citet{parikh2014}. In the following section we will present a proximal gradient descent algorithm for solving \eqref{lasso2} and \eqref{adalasso} in the form of \eqref{lassobig}. \\
To develop some intuition for the proximal gradient algorithm let us first consider the unconstrained minimization of the convex differentiable function $f(\boldsymbol{\theta})$. At the global minimum of $f(\boldsymbol{\theta})$ a necessary and sufficient condition for the optimality of parameters $\boldsymbol{\theta}^{*} \in \mathbb{R}^{p}$ is given by the zero-gradient condition $\nabla f(\boldsymbol{\theta}^{*})=0$.  Typically, gradient descent methods require two primary decisions be made at each successive iteration. First, a direction of descent must be determined.  This direction will be the direction of steepest descent $-\nabla f(\boldsymbol{\theta}^{s})$ for $s=0,1,2,\ldots$.  Second, a step size (or scale factor) must be chosen to govern the size of the step taken. This step size is governed by a step size parameter $\gamma^{s}$, such that  $\boldsymbol{\theta}^{s+1}=\boldsymbol{\theta}^{s}-\gamma^{s}\nabla f(\boldsymbol{\theta}^{s})$ or equivalently 
\begin{equation}
\label{f1}
   \boldsymbol{\theta}^{s+1} = 
   \argmin_{\boldsymbol{\theta} \in \mathbb{R}^{p}} 
  \left\{
  f(\boldsymbol{\theta}^{s}) + \langle \nabla f(\boldsymbol{\theta}^{s}), \boldsymbol{\theta}-\boldsymbol{\theta}^{s} \rangle + \frac{1}{2\gamma^{s}} \| \boldsymbol{\theta}-\boldsymbol{\theta}^{s}\|^{2}_{2}
  \right\}.
\end{equation} 
\noindent Here, we can see the unconstrained minimization problem in \eqref{f1} is simply the local linear approximation to $f(\boldsymbol{\theta})$ supplemented with a quadratic smoothness term.\\
Unlike the problem in \eqref{f1} the optimization problems described in \eqref{lasso2} and \eqref{adalasso} are both nondifferentiable due to presence of the weighted $\ell_{1}$ penalty. At this point it is helpful to consider the decomposition of $f(\boldsymbol{\theta})$ into separable components $g(\boldsymbol{\theta})$ and $h(\boldsymbol{\theta})$ such that $f(\boldsymbol{\theta}) \coloneqq g(\boldsymbol{\theta}) + h(\boldsymbol{\theta})$ where $g(\boldsymbol{\theta})$ is convex and differentiable and $h(\boldsymbol{\theta})$ is convex but nondifferentiable. In doing so we can define a gradient update where $g(\boldsymbol{\theta})$ is approximated as in \eqref{f1} and we leave the nonsmooth $h(\boldsymbol{\theta})$ in its original form
\begin{equation}
\label{f2}
   \boldsymbol{\theta}^{s+1} = 
   \argmin_{\boldsymbol{\theta} \in \mathbb{R}^{p}} 
  \left\{
  g(\boldsymbol{\theta}^{s}) + \langle \nabla g(\boldsymbol{\theta}^{s}), \boldsymbol{\theta}-\boldsymbol{\theta}^{s} \rangle + \frac{1}{2\gamma^{s}} \| \boldsymbol{\theta}-\boldsymbol{\theta}^{s}\|_2^{2}+h(\boldsymbol{\theta})
  \right\}.
\end{equation} 
Now, for the weighted LASSO problem in \eqref{lassobig}, this decompositions takes the form
\begin{align}
\label{proxg}
g(\boldsymbol{\theta}) &= \frac{1}{N} \|\mathscrbf{Y}-\mathscrbf{Z}\boldsymbol{\theta}\|^{2}_{2}\\
\label{proxh}
h(\boldsymbol{\theta}) &=\lambda_{1}\|\boldsymbol{\theta}\|_{1,\mathbf{w}}=\lambda_{1}\sum_{i}w_{i}|\theta_{i}|,
\end{align}
where $\nabla g({\boldsymbol{\theta}})=\mathscrbf{Z}^{'}(\mathscrbf{Y}-\mathscrbf{Z}{\boldsymbol{\theta}})$ and the composition of $\mathbf{w}$ is determined both by the similarity of individuals in the sample and the nature of the penalization scheme.\\

Fortunately, in the case of \eqref{lassobig} the proximal operator for $g(\boldsymbol{\theta})$ has a closed-form solution whose evaluation is negligible in terms of computational costs. We can write the $i$th component of the proximal operator $\mathbf{prox}_{h,\lambda_{1}}$ as
\begin{equation}
\label{prox}
   (\mathbf{prox}_{h,\lambda_{1}}(\boldsymbol{\theta}))_{i} = 
    \mathbf{prox}_{\lambda, w_{i}}(\theta_{i}) = 
    \left\{
  	\begin{array}{@{}ll@{}}
    	\theta_{i} + \lambda_{1} w_{i}, & \text{if}\ \theta_{i} < -\lambda_{1} w_{i} \\
    	0, & \text{if}\ |\theta_{i}| \leq \lambda_{1} w_{i}\\
	\theta_{i} - \lambda_{1} w_{i}, & \text{if}\ \theta_{i} > \lambda_{1} w_{i}\\
  	\end{array} \right.
\end{equation} 
\noindent due to the separable sum property and the definition of the weighted $\ell_{1}$ norm. Using \eqref{prox} we can now write gradient update given \eqref{f2} as a proximal gradient update 
\begin{align}
\label{update1}
\boldsymbol{\theta}^{s+1} & =
	\mathbf{prox}_{h,\gamma^{s}} 
	\left\{
	\boldsymbol{\theta}^{s} - \gamma^{s}\nabla g(\boldsymbol{\theta}^{s}) 
	\right\} \\
\label{update2}
	& = 
	\mathbf{prox}_{h,\gamma^{s}} 
	\left\{
	\boldsymbol{\theta}^{s} - \gamma^{s}\left(\mathscrbf{Z}^{'}(\mathscrbf{Y}-\mathscrbf{Z}{\boldsymbol{\theta}}^{s})\right)
	\right\},
\end{align} 
\noindent where the precomputation of $\mathscrbf{Z}^{'}\mathscrbf{Z}$ and $\mathscrbf{Z}^{'}\mathscrbf{Y}$ can further reduce the computational cost of each update as the objective functional value will only differ by a constant. A classic proximal gradient schemes for solving \eqref{lassobig} is the Iterative Shrinkage-thresholding Algorithm (ISTA). In the standard ISTA formulation step size is treated as a constant across descent iterations and no smoothing techniques are used to accelerate the descent.  To overcome these limitations \citet{beck2009} proposed a general Fast Iterative Shrinkage-Thresholding Algorithm (FISTA) for solving gradient descent problems. In the remainder of this section we describe the version of FISTA we have implemented for the multi-VAR problems described above.\\
As mentioned previously the choice of the step size parameter $\gamma^{s}$ in gradient descent can have a large impact on the convergence rate of the estimator, and also whether a global minimum is reached. One convenient method for determining an approximately optimal step size is to perform a backtracking line search \citep{boyd2004} within each iteration. In this scheme the step size is determined by iteratively rescaling $\gamma$ by $\eta$ where $\eta \in (0,1)$ until 
\begin{equation}
\label{backtrack}
f(\boldsymbol{\theta}-\nabla f(\boldsymbol{\theta})) \leq
f(\boldsymbol{\theta}) - \gamma\alpha \|\nabla f(\boldsymbol{\theta})\|^{2},
\end{equation}
\noindent where $\alpha \in (0,0.5)$ is the second constant, in addition to $\eta$, used to govern the backtracking procedure. Based on previous experience we have chosen a value of $\alpha=0.5$, which corresponds to a maximum decrease in $f$ between $1\%$ and $50\%$ and $\eta=0.5$ which corresponds to a moderate value of granularity as \citet[][p. 466]{boyd2004} suggest $\eta$ should be chosen within the range of $0.1$ (more crude search) and $0.8$ (less crude search).\\
A final improvement to the typical gradient descent procedures corrects the "zig-zagging" descent often observed during iterative computation of \eqref{update1}, which may slow convergence \citep{hastie2015}. A solution initially proposed by \citet{nesterov2007} and incorporated into FISTA by \citet{beck2009} uses weighted combinations of the past gradient descent directions to smooth the global descent path. Another nice feature of proximal gradient descent is that the acceleration approach suggested by \citet{nesterov2007} can be integrated into the proximal operator such that the gradient step now involves
\begin{align}
\label{nesterov}
c^{s+1} & \coloneqq 0.5\left(1+\sqrt{1+4(c^{s})^2}\right)\\
\boldsymbol{\theta}^{s+1} & =
	\mathbf{prox}_{\gamma^{s},h} 
	\left\{
	\boldsymbol{\psi}^{s} - \gamma^{s}\nabla g(\boldsymbol{\psi}^{s}) 
	\right\} \\
\boldsymbol{\psi}^{s+1} & = \boldsymbol{\theta}^{s+1} + \frac{c^{s-1}}{c^{s+1}}(\boldsymbol{\theta}^{s+1}-\boldsymbol{\theta}^{s})
\end{align}
where the step size $\gamma^{s}$ is chosen by the procedure while iterating until \eqref{backtrack} is met. The constant $c^{s}$ is updated at each iteration.  Finally, we provide pseudocode describing our algorithm in full.\\ 
 
\begin{algorithm}[t]
\SetKwRepeat{Do}{do}{while}
\setstretch{1}
\SetAlgoLined
\KwIn{Set $\boldsymbol{\theta}^{0}=\boldsymbol{\psi}^{0}=\mathbf{0}$, $c^0=1$, $\alpha=0.5$, $\eta=0.5$, $\gamma=0.5$, choose an $\epsilon>0$.} 
\KwOut{A solution $\boldsymbol{\theta}^{s}$.}
\For{$s = 0,\dots,s_{max}$}{
Update $c^{s+1} \coloneqq 0.5\left(1+\sqrt{1+4(c^{s})^2}\right)$  and terminate if $\|\boldsymbol{\theta}^{s+1}-\boldsymbol{\theta}^{s}\|_{2}\leq \epsilon\: \text{max}\{1,\|\boldsymbol{\theta}^{s}\|_{2}\}$ \\
\While{ 
$f(\boldsymbol{\theta}-\nabla f(\boldsymbol{\theta})) >
f(\boldsymbol{\theta}) - \gamma\alpha \|\nabla f(\boldsymbol{\theta})\|^{2}$ 
}{
    \begin{equation}
     \begin{cases}
     \boldsymbol{\theta}^{s+1}  =\mathbf{prox}_{\gamma^{s},h} \left\{\boldsymbol{\psi}^{s} - \gamma^{s}\nabla g(\boldsymbol{\psi}^{s}) \right\} \\
     \boldsymbol{\psi}^{s+1} = \boldsymbol{\theta}^{s+1} + \frac{c^{s-1}}{c^{s+1}}(\boldsymbol{\theta}^{s+1}-\boldsymbol{\theta}^{s})\\
     \gamma^{s} =  \eta\gamma^{s}
    \end{cases}
	\end{equation}
  }
 }
 \caption{Fast Iterative Shrinkage-Thresholding Algorithm (FISTA) for Solving \eqref{lassobig}}
\end{algorithm}

\subsection{Forecasting from the Estimated VAR Process}

Here we provide a brief description of how forecasts are obtained from the estimated VAR(1) transition matrices. For the multi-VAR approach $1$-step ahead linear predictions of  $\mathbf{Y}^{(k)}_{T+1}$ for individual $k$ is given by
\begin{equation}
\label{1fcast}
\mathbf{Y}^{(k)}_{T+1} = \mathbf{Z}_{T}^{(k)}(\hat{\boldsymbol{\mu}}+\hat{\boldsymbol{\Delta}}_{k}).
\end{equation}
From \eqref{1fcast} the $h$-step ahead forecast can be computed recursively for any horizon $h$.

\subsection{Selection of the Penalty Parameters}
Performance of the proposed multi-VAR approach is dependent on the selection of the unknown penalty parameters $\lambda_{1}$ and $\lambda_{2, k}$, $k=1,\dots,K$. Here we provide additional details on how the penalty parameters are chosen in the simulation studies and empirical example. The first step of our proposed procedure involves constructing a grid of plausible penalty values. Following \citet{friedman2010} we first identify $\lambda_{1,max}$, or the smallest value of $\lambda_{1}$ for which all the coefficients in the model will be zero. In the multi-VAR setting $\lambda_{1,max}$ is equal to $\textrm{max}|\mathscrbf{Z}^{'}\mathscrbf{Y}|$ where $\mathscrbf{Z}$ and $\mathscrbf{Y}$ are given in \eqref{lassobig}. From $\lambda_{1,max}$ we construct a grid from $\lambda_{1,max}$ to $\lambda_{1,max}/1000$ using equally spaced values on a log-scale. Across a number of data contexts $20$ values of $\lambda_{1}$ and $\lambda_{2, k}$ were found to be sufficient. Following \citet[p. 32]{ollier2014} the ratio $\lambda_{2}/\lambda_{1,k}$ is chosen to vary on the interval $(0,\dots,K)$ and this ratio is used to solve for $\lambda_{2,k,max}$, from which another grid is constructed, $\lambda_{2,k,max}$ to $\lambda_{2,k,max}/1000$, also on a log-linear scale. 

To identify the optimal penalty parameters from our grid of candidate values we adapt the rolling-window cross-validation (RWCV) procedure for high-dimensional VAR models described by \citet{banbura2010}, \citet{song2011}, and \citet{nicholson2017} to the multi-VAR problem. This procedure involves searching across the grid of predetermined values described above and choosing the combination of penalty parameters that minimize the $h$-step ahead mean-square forecast error (MSFE). Here $h=1,2,3,\dots$ is the desired forecast horizon and for all analyses in this paper $h=1$ is used to select the penalty parameters.  We chose to use this forecast horizon as ILD is often collected daily and we hypothesized that one-day-ahead predictions have utility in behavioral, health and social science applications. To implement the rolling-window cross-validation procedure we divide each individual dataset into three periods. The first period is the initialization period beginning at the first time point and ending at $T_{1}$. Based on the literature above we set $T_{1}=T/3$, or approximately $1/3$ of the time series length. The second period is the training period, starting at $T _{1}+1$ and ending at $T_{2}$. We chose $T_{2}$ to be equal to $T-3$, leaving a hold-out-sample of three observations in the final period ($T_{2}+1$ to $T$) for our pseudo-out-of-sample forecast evaluation.

For each value of the $\lambda_1$ and $\lambda_{2,k}$ grid we perform the following sequence. First, we solve the problem in \eqref{lassobig} using timepoints $1,\dots,T_{1}$ from each individual in the sample to obtain $\hat{\boldsymbol{\mu}}$ and $\hat{\boldsymbol{\Delta}}_{k}$. Separately for each individual these estimates are then used to forecast $\hat{\mathbf{Y}}^{(k)}_{T_{1}+1}$ and obtain the MSFE. We continue in this fashion, adding one observation at a time to the initialization period and forecasting ahead $h$-units until we reach $T_{2}-h$, at which time the forecast performance is aggregated across the $(T_{2}-T_{1}-h+1)$ forecasts  for each combination of $\lambda_{1}$ and $\lambda_{2,k}$ as in
\begin{equation}
\text{MSFE}^{(k)}_{\lambda_{1},\lambda_{2,k}}= \frac{1}{T_{2}-T_{1}+1}
\sum^{T_{2}-1}_{t=T_{1}}\| \hat{\mathbf{Y}}^{(k)}_{t+1}-\mathbf{Y}^{(k)}_{t+1} \|^{2}_{2},
\end{equation}
and the values of $\lambda_{1}$ and $\lambda_{2,k}$ which correspond to the smallest MSFE are chosen for evaluating the forecast performance in the hold-out sample.

\section{Performance Evaluation}
To better understand the finite sample properties of the proposed models and algorithm we conduct a Monte Carlo simulation designed to replicate some of the basic features of ILD collected from multiple subjects. 

\subsection{Simulation Design}
To evaluate the performance of the proposed approach for forecasting multiple subject ILD we generated data according to a number of commonly encountered features: (1) individual time series lengths of $T=(30, 50, 100)$, (2) number of ILD variables collected per individual $d=(10,20, 30)$, (3) total number of individuals in the sample $K=(10, 20)$, (4) the level of cross-sectional heterogeneity (low, medium and high) and (5) the type of penalized VAR model employed; (a) VAR fit by LASSO for each individual in the sample separately as in \eqref{lasso}, (b) the multi-VAR as in \eqref{lasso2}, and (c) the adaptive multi-VAR(1) as in \eqref{adalasso}.\\
Across all design factors the $d \times d$ sparse transition matrices for each individual were generated to have $5\%$ nonzero entries. This means, for example, a multivariate time series with $d=30$ would have 45 nonzero coefficients in the data generating model. The position of non-zero elements in each individual's transition matrix were selected randomly given the following constraints. In the low-heterogeneity condition, $2/3$ of paths were common to all individuals, and $1/3$ of paths were completely unique to each individual. In the medium heterogeneity condition, $1/2$ of each individual's paths were common and $1/2$ were unique. In the high-heterogeneity condition $1/3$ were common and $2/3$ were unique. Across all conditions the individual transition matrix elements were drawn from $\mathcal{U}(0.1, 0.9)$ until the stability condition from \eqref{varp} was satisfied. For each of the $2 \times 3 \times 3 \times 3$ data generating conditions we conducted 20 replications.  \\
Across all conditions the $3$ final time points of each component series were withheld to evaluate forecast accuracy. For the synthetic data examples cross-validation was performed in two different ways. First, we assumed the non-zero transition matrices were known and chose the penalty parameters that resulted in the smallest estimation error (e.g. $\|\hat{\mathbf{B}}-\mathbf{B}\|_{F}/\|\mathbf{B}\|_{F})$. This allowed us to compare the different approaches independent of the cross-validation method. Of course, in real data scenarios $\mathbf{B}$ is unknown and it is important to examine our proposed framework under realistic conditions. To this end, our second approach used the RWCV procedure described earlier to select optimal values of $\lambda_{1}$ and $\lambda_{2}$ in our simulation study.
\subsection{Outcome Measures} 
To evaluate the performance of our approach we looked at a number of measures relevant to forecast performance.  These measures include (a) sensitivity, (b) specificity, and (c) root mean square forecast error (RMSFE). The mean sensitivity and the mean specificity were calculated as
\begin{align}
\label{sensitivity}
\text{Mean sensitivity} &= \frac{1}{K}\sum^{K}_{k=1}\left(\frac{\Sigma_{j}(\hat{B}_{k,j} \neq 0 \:\text{and} \:{B}_{k,j} \neq 0)}{\Sigma_{j} ({B}_{k,j} \neq 0)}\right),\\
\label{specificity}
\text{Mean specificity} &= \frac{1}{K}\sum^{K}_{k=1}\left(\frac{\Sigma_{j} (\hat{B}_{k,j} = 0 \:\text{and} \:{B}_{k,j} = 0)}{\Sigma_{j} ({B}_{k,j} = 0)}\right)
\end{align}
where ${B}_{k,j}$ and $\hat{B}_{k,j}$ are the true and the estimated transition matrix elements, respectively, for individual $k$ in a given design condition. Finally, the mean RMSFE is
\begin{equation}
\text{Mean RMSFE} = 
\frac{1}{K}\sum^{K}_{k=1}
\sqrt{\frac{1}{d}\sum^{d}_{j=1}(\hat{\mathbf{Y}}^{(k)}_{j,t-3+h}-\mathbf{Y}^{(k)}_{j,t-3+h})^{2}}
\end{equation}
where $(\hat{\mathbf{Y}}^{(k)}_{j,t-3+h}-\mathbf{Y}^{(k)}_{j,t-3+h})$ is the $h$ step ahead forecast error for individual $k$ on variable $j$ and $h \in \{1,2,3\}$.

\subsection{Simulation Results}

\begin{figure}[h]
\caption{Sensitivity and Specificity for $\mathbf{B}_{k}$ (Estimation Error Approach)}
\label{fig:a}
\centering
\includegraphics[width=1.05\textwidth]{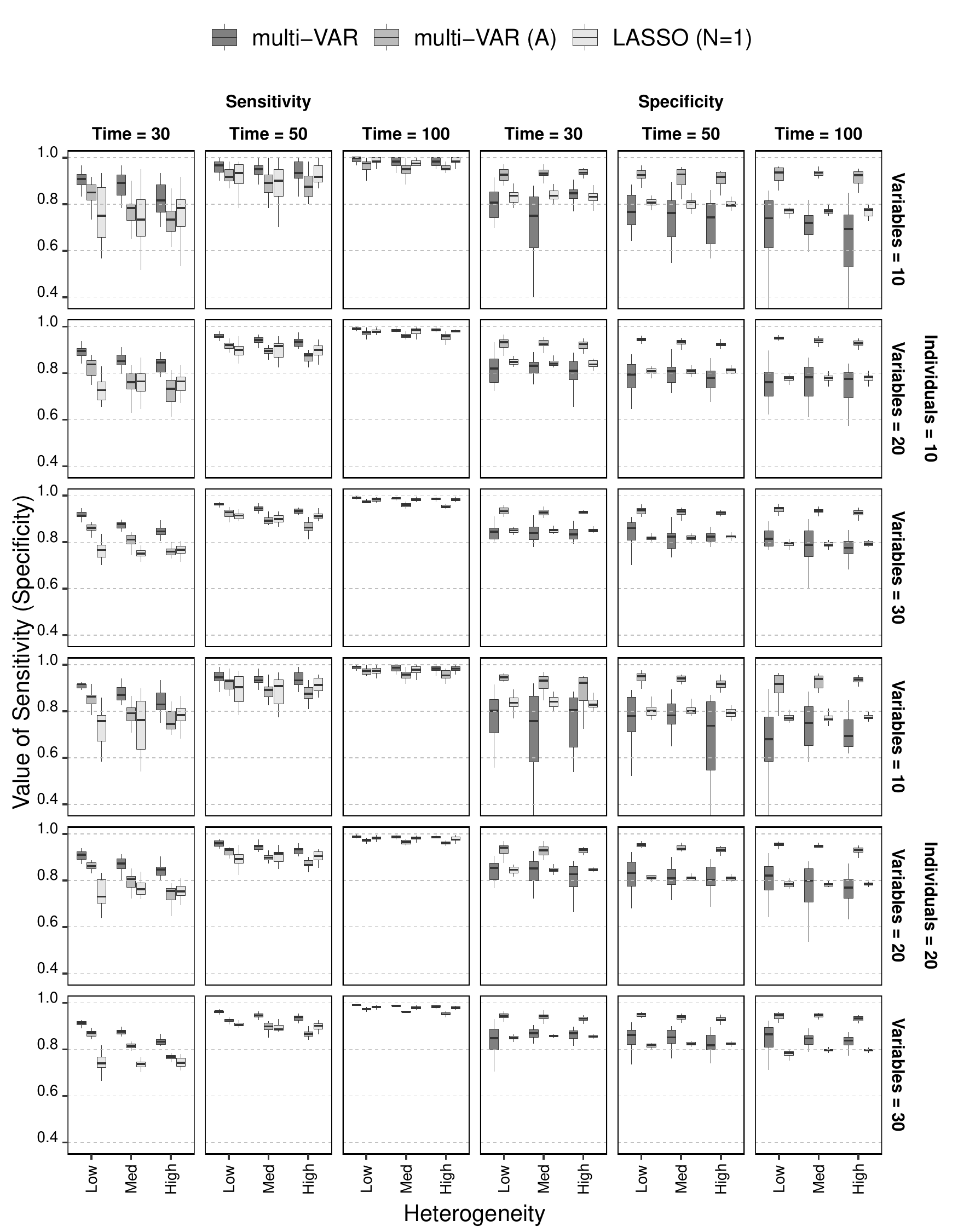}
\end{figure}

\begin{figure}[h]
\caption{Sensitivity and Specificity for $\mathbf{B}_{k}$ (Rolling Window Cross-Validation)}
\label{fig:b}
\centering
\includegraphics[width=1.05\textwidth]{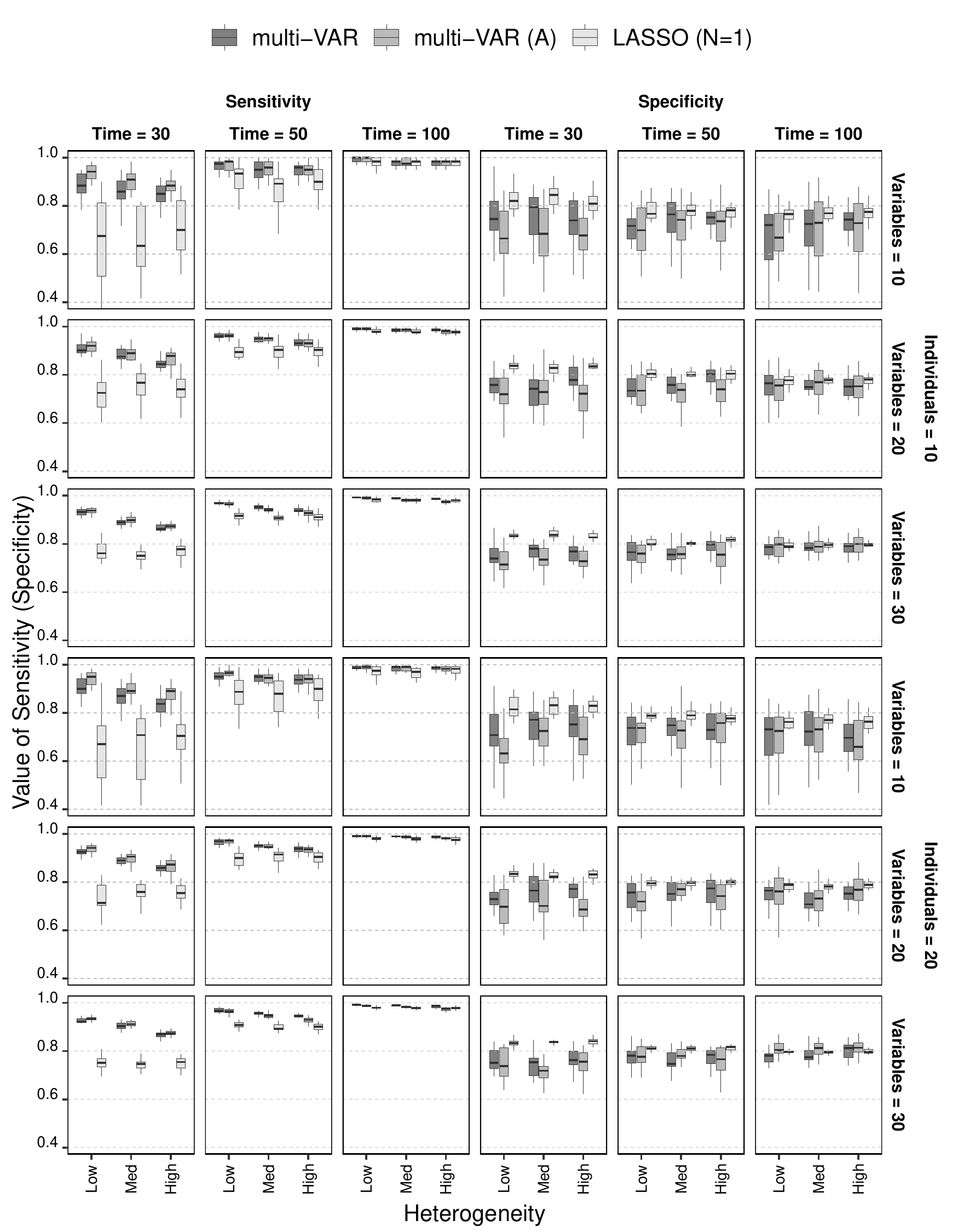}
\end{figure}

As expected performance differences emerged when the penalty parameters were selected using the estimation error metric compared to the rolling-window cross-validation. This is clear when one compares Figures \ref{fig:a} and \ref{fig:b}. An interesting pattern also emerges when comparing the performance of each method in aggregate. For example, the individual LASSO achieved a mean sensitivity of $0.88$ and a mean specificity of $0.81$ across all conditions when the true data generating matrices were used to select $\lambda_1$ and $\lambda_2$, using the estimation error metric. When the RWCV procedure was used mean sensitivity was $0.87$ and a mean specificity $0.80$. The standard multi-VAR achieved a mean sensitivity of $0.93$ and a mean specificity of $0.78$ for the estimation error condition and a sensitivity of $0.94$ and specificity of $0.75$ using RWCV. In aggregate, for the individual-level LASSO and standard multi-VAR approaches sensitivity and specificity were similar across the two penalty selection procedures. On the other hand, performance of the adaptive multi-VAR approach differed considerably across the two approaches, with sensitivity increasing from $0.88$ to $0.94$ and specificity decreasing from $0.93$ to $0.73$. This suggests the adaptive multi-VAR approach suffered the most from the RWCV approach to selecting the penalty parameters. Regardless, the RWCV approach is what will be used in practice we will focus our discussion on this set of results.\\

\subsubsection{Sensitivity and Specificity for $\mathbf{B}_{k}$}

We first consider recovery of the total effect matrix for each individual. Although we would not expect the heterogeneity levels or number of individuals in the sample to impact the individual LASSO performance, we are certainly interested in their impact on the multi-VAR methods. Overall, the differing levels of heterogeneity had little impact on parameter recovery outside of small decrement in sensitivity for the multi-VAR approaches at the smallest time series length of $T=30$. This is also consistent with the aggregate findings as the multi-VAR approaches both showed a decrement ($0.03$) in sensitivity and a slight increase ($0.01-0.02$) in specificity, as heterogeneity increased from low to high. In aggregate, both multi-VAR approaches showed no change in sensitivity or specificity when the number of individuals included in the sample was $10$ or $20$.\\
In contrast to the cross-sectional heterogeneity and the number of individuals conditions, we would expect the number of timepoints and the number of variables to impact the performance of all three approaches. For the standard multi-VAR sensitivity increased as the time series length increased, from $(0.88, 0.95, 0.99)$ for $(T=30,50,100)$, respectively, while specificity remained relatively constant, $(0.75,0.74.0.74)$. For the multi-VAR approach both sensitivity and specificity increased as the time series length increased. Sensitivity and specificity ranged from ($0.90,0.95,0.98$) and ($0.70,0.74,0.75$), respectively. For the individual-level LASSO sensitivity also increased as the time series length increased, $(0.73, 0.90, 0.98)$, while specificity decreased with larger sample sizes $(0.83, 0.80, 0.78)$.\\
The standard multi-VAR showed slight increases in both specificity as the number of variables included in the analysis increased from $10$ to $30$. Specificity and sensitivity increased by $0.02$ and $0.05$, respectively. Similarly, the individual LASSO showed increases in sensitivity $0.04$ and specificity $0.02$ as the number of included variable increased. The multi-VAR approach showed no changes in sensitivity as the number of variables increased, and a small increase in specificity, from $0.69$ to $0.76$. It should be noted that for the individual-LASSO the time series lengths and variable dimensions considered here are quite small.
   
\subsubsection{Sensitivity and Specificity for $\boldsymbol{\mu}$}

\begin{figure}[h]
\caption{Sensitivity and Specificity for $\boldsymbol{\mu}$ (Rolling Window Cross-Validation)}
\label{fig:c}
\centering
\includegraphics[width=1.05\textwidth]{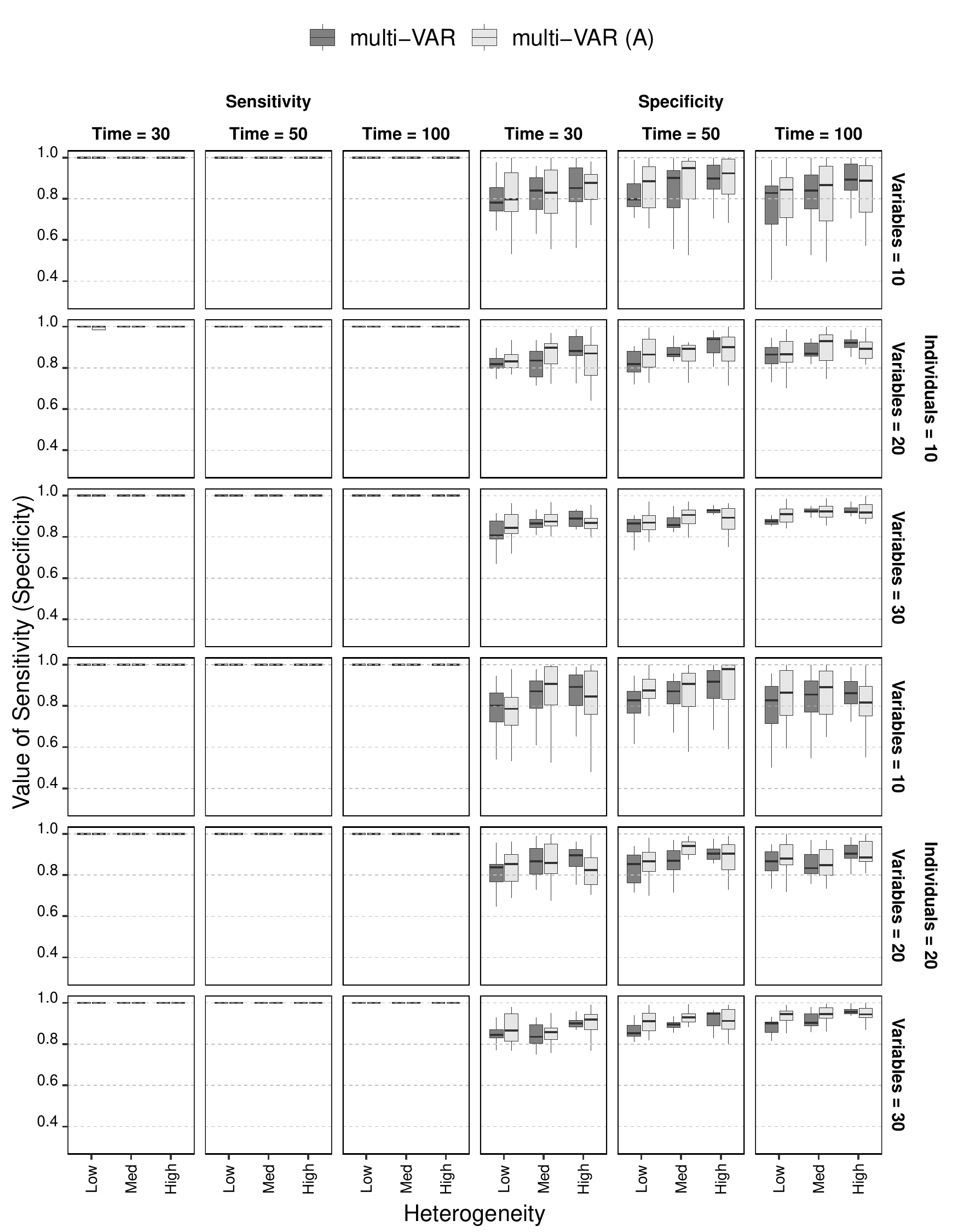}
\end{figure}

In addition to looking at the recovery of $\mathbf{B}_{k}$, we are also interested in the recovery of the common effect matrix $\boldsymbol{\mu}$.  Figure \ref{fig:c} shows the sensitivity and specificity for the two multi-VAR approaches in recovering the common effects. We do not include the individual-level LASSO here as it does not explicitly model group and unique model components. It is clear from the sensitivity plots in Figure \ref{fig:c} that both multi-VAR procedures do well in consistently capturing the common effects across all simulation blocks (sensitivity = $0.99$). In terms of specificity, averaged over all simulation conditions, the standard multi-VAR obtains a specificity of $0.86$ and the adaptive version $0.87$. Importantly, although we see slight increases in specificity as heterogeneity increases and $\boldsymbol{\mu}$ becomes sparser, both approaches perform well in terms of identifying the zero elements of $\boldsymbol{\mu}$. 

\begin{figure}[h]
\caption{Sensitivity and Specificity for $\boldsymbol{\Delta}_{k}$ (Rolling Window Cross-Validation)}
\label{fig:d}
\centering
\includegraphics[width=1.05\textwidth]{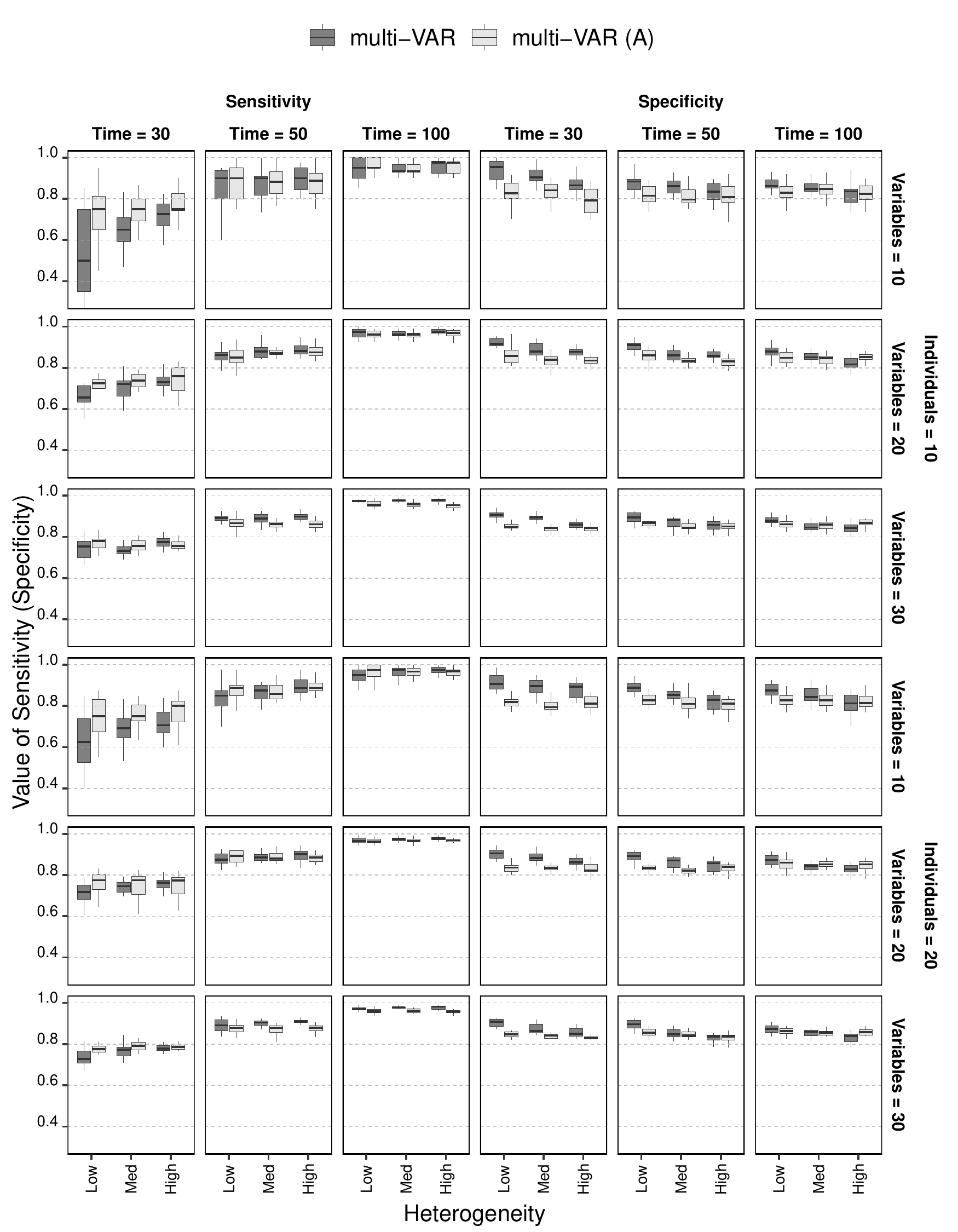}
\end{figure}

It is also worth examining the recovery of the unique effect matrices $\boldsymbol{\Delta}_{k}$.  Figure \ref{fig:d} shows the sensitivity and specificity for the two multi-VAR approaches in recovering the effects unique to each individual. As one might expect the sensitivity of our approaches for recovering $\boldsymbol{\Delta}_{k}$ is lower when compared to the common effect matrix, as these effects are less persistent in the parameter space. Here we see a slight increase in sensitivity to recovering $\boldsymbol{\Delta}_{k}$ as $\boldsymbol{\mu}$ becomes sparser, but this is mostly at the smallest number of timepoints $T=30$. Specificity of both approaches remains high across all simulation conditions suggesting that recovery of the unique effects is quite balanced across the two performance measures.

\subsubsection{Forecasting Performance}

The RMSFE across all simulation conditions and forecast horizons are presented in Table \ref{tab1}. Again, these results are tabulated using the penalty parameters chosen with RWCW procedure. As mentioned previously, for the individual LASSO we would not expect forecasting performance to be impacted by heterogeneity levels or the number of individuals in the sample. In aggregate, collapsing across the other simulation factors, this was also true for both multi-VAR approaches. In addition, for both time series length and the number of variables, all three penalized regression procedures showed small decreases in RMSFE for the 1-unit forecast horizon as time series length and number of variables increased. For forecast horizons of $2$ and $3$ there was less variability across the approaches and simulation factors.

For the simulation conditions we also examined the performance of various benchmark methods; (a) the series mean, (b) the AR(1) model for each component series, and (c) the VAR(1) model. The RMSFE for these benchmark methods are presented in Table \ref{tab2}. For the  forecast horizon of $1$ the regularization methods outperformed the benchmark methods in terms of RMSFE, (LASSO = $0.89$, multi-VAR = $0.87$, multi-VAR (A) = $0.89$, mean = $1.00$,  AR(1) = $1.00$, VAR(1) = 1.26). For the  forecast horizon of $2$, the estimators performed more similarly, (LASSO = $0.95$, multi-VAR = $0.95$, multi-VAR (A) = $0.95$, mean = $1.00$,  AR(1) = $1.00$, VAR(1) = 1.31). This trend continued for the forecast window of $3$, (LASSO = $0.97$, multi-VAR = $0.97$, multi-VAR (A) = $0.97$, mean = $1.00$,  AR(1) = $1.00$, VAR(1) = 1.34).
   
\begin{table}[ht]
\centering
\caption{Root Mean Squared Forecast Error for Simulation Study Conditions} 
\label{tab1}
\scalebox{0.61}{
\begin{threeparttable}
\begin{tabular}{cccccccccccc}
  \toprule
  & & &    \multicolumn{9}{c}{H Step-Ahead Forecast}  \\
 \cmidrule(lr){4-12} & & &  \multicolumn{3}{c}{H=1} &
        \multicolumn{3}{c}{H=2} &
        \multicolumn{3}{c}{H=3} \\
 \cmidrule(lr){4-6}\cmidrule(lr){7-9}\cmidrule(lr){10-12}      \multicolumn{3}{c}{Number of} &
        \multicolumn{3}{c}{Model} &
        \multicolumn{3}{c}{Model} &
        \multicolumn{3}{c}{Model} \\
 \cmidrule(lr){1-3}\cmidrule(lr){4-6}\cmidrule(lr){7-9}\cmidrule(lr){10-12}     \multicolumn{1}{c}{Subjects} &
        \multicolumn{1}{c}{\small Variables} &
        \multicolumn{1}{c}{\small Time} &
        \multicolumn{1}{c}{\small LASSO} &
        \multicolumn{1}{c}{\small m-VAR} &
        \multicolumn{1}{c}{\small m-VAR (A)} &
        \multicolumn{1}{c}{\small LASSO} &
        \multicolumn{1}{c}{\small m-VAR} &
        \multicolumn{1}{c}{\small m-VAR (A)} &
        \multicolumn{1}{c}{\small LASSO} &
        \multicolumn{1}{c}{\small m-VAR} &
        \multicolumn{1}{c}{\small m-VAR (A)}\\
   \arrayrulecolor{black}\midrule
  \multicolumn{12}{c}{\textbf{Heterogeneity: Low}}\\
   \arrayrulecolor{black}\midrule
\multirow{9}{*}{10} & \multirow{3}{*}{10} & 30 & 0.96 & 0.97 & 0.98 & 0.94 & 0.94 & 0.94 & 0.98 & 0.99 & 1.00 \\ 
  & & 50 & 0.94 & 0.94 & 0.94 & 0.97 & 0.96 & 0.95 & 0.98 & 0.99 & 0.99 \\ 
  & & 100 & 0.96 & 0.96 & 0.94 & 0.96 & 0.95 & 0.95 & 1.00 & 0.98 & 1.00 \\ 
   \arrayrulecolor{black!50}\cmidrule(lr){2-12} 
& \multirow{3}{*}{20} & 30 & 0.87 & 0.89 & 0.91 & 0.93 & 0.94 & 0.94 & 0.98 & 0.99 & 0.98 \\ 
  &  & 50 & 0.85 & 0.87 & 0.89 & 0.97 & 0.97 & 0.97 & 0.99 & 0.99 & 0.97 \\ 
  &  & 100 & 0.86 & 0.87 & 0.88 & 0.94 & 0.94 & 0.97 & 0.97 & 0.97 & 0.98 \\ 
   \arrayrulecolor{black!50}\cmidrule(lr){2-12} 
& \multirow{3}{*}{30}& 30 & 0.85 & 0.88 & 0.89 & 0.96 & 0.98 & 0.98 & 0.99 & 1.00 & 1.00 \\ 
  &  & 50 & 0.81 & 0.82 & 0.83 & 0.93 & 0.93 & 0.96 & 0.98 & 0.98 & 0.98 \\ 
  &  & 100 & 0.79 & 0.80 & 0.79 & 0.91 & 0.91 & 0.89 & 0.95 & 0.95 & 0.95 \\ 
   \arrayrulecolor{black!50}\cmidrule(lr){1-12} 
\multirow{9}{*}{20} & \multirow{3}{*}{10} & 30 & 0.91 & 0.92 & 0.94 & 0.95 & 0.95 & 0.95 & 0.96 & 0.96 & 0.96 \\ 
  &  & 50 & 0.92 & 0.93 & 0.91 & 0.97 & 0.97 & 0.95 & 0.97 & 0.98 & 0.96 \\ 
  & & 100 & 0.92 & 0.92 & 0.93 & 0.95 & 0.96 & 0.95 & 0.98 & 0.98 & 0.96 \\ 
   \arrayrulecolor{black!50}\cmidrule(lr){2-12} 
& \multirow{3}{*}{20} & 30 & 0.89 & 0.92 & 0.93 & 0.94 & 0.95 & 0.95 & 0.98 & 0.98 & 0.98 \\ 
  &  & 50 & 0.86 & 0.87 & 0.88 & 0.93 & 0.94 & 0.95 & 0.95 & 0.95 & 0.96 \\ 
 &  & 100 & 0.85 & 0.85 & 0.85 & 0.95 & 0.95 & 0.96 & 0.96 & 0.96 & 0.98 \\ 
   \arrayrulecolor{black!50}\cmidrule(lr){2-12} 
& \multirow{3}{*}{30} & 30 & 0.84 & 0.87 & 0.89 & 0.94 & 0.96 & 0.96 & 0.96 & 0.97 & 0.97 \\ 
   &  & 50 & 0.80 & 0.81 & 0.84 & 0.91 & 0.92 & 0.91 & 0.94 & 0.94 & 0.96 \\ 
  &  & 100 & 0.81 & 0.82 & 0.83 & 0.92 & 0.92 & 0.92 & 0.97 & 0.97 & 0.97 \\ 
   \arrayrulecolor{black}\midrule
  \multicolumn{12}{c}{\textbf{Heterogeneity: Medium}}\\
   \arrayrulecolor{black}\midrule
\multirow{9}{*}{10} & \multirow{3}{*}{10}  & 30 & 0.91 & 0.93 & 0.93 & 0.96 & 0.96 & 0.96 & 0.93 & 0.93 & 0.94 \\ 
  &  & 50 & 0.93 & 0.94 & 0.91 & 0.99 & 0.99 & 0.98 & 0.99 & 0.99 & 1.02 \\ 
  &  & 100 & 0.92 & 0.92 & 0.91 & 0.99 & 0.99 & 0.96 & 0.95 & 0.94 & 0.94 \\ 
        \arrayrulecolor{black!50}\cmidrule(lr){2-12} 
  & \multirow{3}{*}{20} & 30 & 0.90 & 0.92 & 0.91 & 0.96 & 0.97 & 0.96 & 0.98 & 0.98 & 0.98 \\ 
  &  & 50 & 0.88 & 0.90 & 0.89 & 0.94 & 0.95 & 0.95 & 0.98 & 0.98 & 0.97 \\ 
   &  & 100 & 0.84 & 0.85 & 0.89 & 0.94 & 0.94 & 0.95 & 0.96 & 0.96 & 0.98 \\ 
                \arrayrulecolor{black!50}\cmidrule(lr){2-12} 
  & \multirow{3}{*}{30} & 30 & 0.85 & 0.88 & 0.90 & 0.97 & 0.99 & 0.98 & 0.97 & 0.98 & 0.98 \\ 
  &  & 50 & 0.83 & 0.85 & 0.83 & 0.94 & 0.95 & 0.94 & 0.99 & 1.00 & 0.97 \\ 
  &  & 100 & 0.79 & 0.80 & 0.80 & 0.88 & 0.88 & 0.91 & 0.94 & 0.94 & 0.95 \\ 
                        \arrayrulecolor{black!50}\cmidrule(lr){1-12} 
  \multirow{9}{*}{20} & \multirow{3}{*}{10}  & 30 & 0.95 & 0.97 & 0.96 & 0.96 & 0.97 & 0.98 & 0.94 & 0.94 & 0.95 \\ 
  &  & 50 & 0.94 & 0.95 & 0.90 & 0.96 & 0.95 & 0.95 & 0.98 & 0.98 & 0.96 \\ 
  &  & 100 & 0.93 & 0.93 & 0.94 & 0.97 & 0.97 & 0.95 & 0.96 & 0.96 & 0.96 \\ 
                                \arrayrulecolor{black!50}\cmidrule(lr){2-12} 
  & \multirow{3}{*}{20} & 30 & 0.88 & 0.91 & 0.91 & 0.94 & 0.95 & 0.95 & 0.97 & 0.97 & 0.97 \\ 
  &  & 50 & 0.87 & 0.88 & 0.88 & 0.94 & 0.95 & 0.95 & 0.98 & 0.98 & 0.99 \\ 
  &  & 100 & 0.86 & 0.86 & 0.87 & 0.97 & 0.98 & 0.97 & 0.99 & 0.99 & 0.99 \\ 
                                        \arrayrulecolor{black!50}\cmidrule(lr){2-12} 
  & \multirow{3}{*}{30} & 30 & 0.86 & 0.88 & 0.88 & 0.94 & 0.95 & 0.95 & 0.97 & 0.98 & 0.98 \\ 
  &  & 50 & 0.82 & 0.84 & 0.84 & 0.93 & 0.94 & 0.95 & 0.97 & 0.97 & 0.97 \\ 
  &  & 100 & 0.81 & 0.82 & 0.81 & 0.92 & 0.93 & 0.93 & 0.97 & 0.97 & 0.97 \\ 
  \arrayrulecolor{black}\midrule
  \multicolumn{12}{c}{\textbf{Heterogeneity: High}}\\
   \arrayrulecolor{black}\midrule
\multirow{9}{*}{10} & \multirow{3}{*}{10} & 30 & 0.92 & 0.94 & 0.93 & 0.97 & 0.97 & 0.97 & 0.94 & 0.94 & 0.93 \\ 
  & & 50 & 0.91 & 0.92 & 0.95 & 0.95 & 0.95 & 0.94 & 0.95 & 0.95 & 0.95 \\ 
  & & 100 & 0.92 & 0.92 & 0.92 & 0.95 & 0.95 & 0.99 & 0.94 & 0.95 & 0.94 \\ 
                                        \arrayrulecolor{black!50}\cmidrule(lr){2-12} 
  & \multirow{3}{*}{20}  & 30 & 0.89 & 0.92 & 0.90 & 0.98 & 0.99 & 0.97 & 0.93 & 0.94 & 0.93 \\ 
  & & 50 & 0.91 & 0.94 & 0.89 & 0.95 & 0.96 & 0.95 & 1.00 & 1.00 & 0.98 \\ 
  &  & 100 & 0.88 & 0.88 & 0.86 & 0.94 & 0.94 & 0.93 & 0.99 & 0.99 & 0.94 \\ 
                                                                                \arrayrulecolor{black!50}\cmidrule(lr){2-12} 
  & \multirow{3}{*}{30}  & 30 & 0.84 & 0.87 & 0.88 & 0.93 & 0.94 & 0.94 & 0.97 & 0.97 & 0.98 \\ 
   & & 50 & 0.81 & 0.83 & 0.82 & 0.91 & 0.92 & 0.93 & 0.97 & 0.97 & 0.97 \\ 
   & & 100 & 0.79 & 0.79 & 0.80 & 0.91 & 0.91 & 0.92 & 0.98 & 0.97 & 0.97 \\ 
                                                      \arrayrulecolor{black!50}\cmidrule(lr){1-12} 
\multirow{9}{*}{20} & \multirow{3}{*}{10}  & 30 & 0.89 & 0.91 & 0.90 & 0.94 & 0.94 & 0.93 & 0.95 & 0.95 & 0.96 \\ 
   &  & 50 & 0.91 & 0.91 & 0.89 & 0.95 & 0.95 & 0.97 & 0.97 & 0.97 & 0.98 \\ 
   &  & 100 & 0.90 & 0.90 & 0.91 & 0.93 & 0.93 & 0.96 & 0.97 & 0.97 & 0.98 \\ 
                                                      \arrayrulecolor{black!50}\cmidrule(lr){2-12} 
  & \multirow{3}{*}{20}  & 30 & 0.90 & 0.93 & 0.92 & 0.95 & 0.96 & 0.96 & 0.98 & 0.98 & 0.98 \\ 
   &  & 50 & 0.87 & 0.89 & 0.87 & 0.96 & 0.97 & 0.97 & 0.99 & 1.00 & 1.00 \\ 
   &  & 100 & 0.88 & 0.88 & 0.89 & 0.95 & 0.95 & 0.95 & 0.98 & 0.98 & 0.98 \\ 
                  \arrayrulecolor{black!50}\cmidrule(lr){2-12} 
  & \multirow{3}{*}{30} & 30 & 0.87 & 0.90 & 0.89 & 0.95 & 0.96 & 0.95 & 0.97 & 0.97 & 0.97 \\ 
   &  & 50 & 0.84 & 0.86 & 0.84 & 0.92 & 0.93 & 0.94 & 0.96 & 0.97 & 0.97 \\ 
   &  & 100 & 0.80 & 0.80 & 0.82 & 0.92 & 0.92 & 0.92 & 0.97 & 0.98 & 0.96 \\ 
   \arrayrulecolor{black}\bottomrule
  \end{tabular}
  \begin{tablenotes}
   \item Note. (A) indicates adaptive version of the multi-VAR.
 \end{tablenotes}
\end{threeparttable}
}
\end{table}

\begin{table}[ht]
\centering
\caption{Root Mean Squared Forecast Error for Benchmark Methods Across Simulation Conditions} 
\label{tab2}
\scalebox{0.8}{
\begin{threeparttable}
\begin{tabular}{cccccccccccc}
  \toprule
  & & &    \multicolumn{9}{c}{H Step-Ahead Forecast}  \\
 \cmidrule(lr){4-12} & & &  \multicolumn{3}{c}{H=1} &
        \multicolumn{3}{c}{H=2} &
        \multicolumn{3}{c}{H=3} \\
 \cmidrule(lr){4-6}\cmidrule(lr){7-9}\cmidrule(lr){10-12}      \multicolumn{3}{c}{Number of} &
        \multicolumn{3}{c}{Model} &
        \multicolumn{3}{c}{Model} &
        \multicolumn{3}{c}{Model} \\
 \cmidrule(lr){1-3}\cmidrule(lr){4-6}\cmidrule(lr){7-9}\cmidrule(lr){10-12}     \multicolumn{1}{c}{Subjects} &
        \multicolumn{1}{c}{\small Variables} &
        \multicolumn{1}{c}{\small Time} &
        \multicolumn{1}{c}{\small Mean} &
        \multicolumn{1}{c}{\small AR(1)} &
        \multicolumn{1}{c}{\small VAR(1)} &
        \multicolumn{1}{c}{\small Mean} &
        \multicolumn{1}{c}{\small AR(1)} &
        \multicolumn{1}{c}{\small VAR(1)} &
        \multicolumn{1}{c}{\small Mean} &
        \multicolumn{1}{c}{\small AR(1)} &
        \multicolumn{1}{c}{\small VAR(1)}\\
   \arrayrulecolor{black}\midrule
\multirow{9}{*}{10} & \multirow{3}{*}{10}  & 30 & 0.92 & 0.94 & 0.93 & 0.97 & 0.97 & 0.97 & 0.94 & 0.94 & 0.93 \\ 
  &  & 50 & 0.91 & 0.92 & 0.95 & 0.95 & 0.95 & 0.94 & 0.95 & 0.95 & 0.95 \\ 
  &  & 100 & 0.92 & 0.92 & 0.92 & 0.95 & 0.95 & 0.99 & 0.94 & 0.95 & 0.94 \\ 
        \arrayrulecolor{black!50}\cmidrule(lr){2-12} 
  & \multirow{3}{*}{20} & 30 & 0.89 & 0.92 & 0.90 & 0.98 & 0.99 & 0.97 & 0.93 & 0.94 & 0.93 \\ 
   &  & 50 & 0.91 & 0.94 & 0.89 & 0.95 & 0.96 & 0.95 & 1.00 & 1.00 & 0.98 \\ 
   &  & 100 & 0.88 & 0.88 & 0.86 & 0.94 & 0.94 & 0.93 & 0.99 & 0.99 & 0.94 \\ 
                \arrayrulecolor{black!50}\cmidrule(lr){2-12} 
  & \multirow{3}{*}{30}& 30 & 0.84 & 0.87 & 0.88 & 0.93 & 0.94 & 0.94 & 0.97 & 0.97 & 0.98 \\ 
   &  & 50 & 0.81 & 0.83 & 0.82 & 0.91 & 0.92 & 0.93 & 0.97 & 0.97 & 0.97 \\ 
   &  & 100 & 0.79 & 0.79 & 0.80 & 0.91 & 0.91 & 0.92 & 0.98 & 0.97 & 0.97 \\ 
                        \arrayrulecolor{black!50}\cmidrule(lr){1-12} 
 \multirow{9}{*}{20} & \multirow{3}{*}{10} & 30 & 0.89 & 0.91 & 0.90 & 0.94 & 0.94 & 0.93 & 0.95 & 0.95 & 0.96 \\ 
   &  & 50 & 0.91 & 0.91 & 0.89 & 0.95 & 0.95 & 0.97 & 0.97 & 0.97 & 0.98 \\ 
   &  & 100 & 0.90 & 0.90 & 0.91 & 0.93 & 0.93 & 0.96 & 0.97 & 0.97 & 0.98 \\ 
                                \arrayrulecolor{black!50}\cmidrule(lr){2-12} 
  & \multirow{3}{*}{20}& 30 & 0.90 & 0.93 & 0.92 & 0.95 & 0.96 & 0.96 & 0.98 & 0.98 & 0.98 \\ 
   &  & 50 & 0.87 & 0.89 & 0.87 & 0.96 & 0.97 & 0.97 & 0.99 & 1.00 & 1.00 \\ 
   &  & 100 & 0.88 & 0.88 & 0.89 & 0.95 & 0.95 & 0.95 & 0.98 & 0.98 & 0.98 \\ 
                                        \arrayrulecolor{black!50}\cmidrule(lr){2-12} 
  & \multirow{3}{*}{30} & 30 & 0.87 & 0.90 & 0.89 & 0.95 & 0.96 & 0.95 & 0.97 & 0.97 & 0.97 \\ 
   &  & 50 & 0.84 & 0.86 & 0.84 & 0.92 & 0.93 & 0.94 & 0.96 & 0.97 & 0.97 \\ 
   &  & 100 & 0.80 & 0.80 & 0.82 & 0.92 & 0.92 & 0.92 & 0.97 & 0.98 & 0.96 \\ 
   \arrayrulecolor{black}\bottomrule
  \end{tabular}
  \begin{tablenotes}
 \end{tablenotes}
\end{threeparttable}
}
\end{table}

\section{An Illustrative Example}

We now present an empirical example based on \citet{fredrickson2017} who examined the day-to-day emotional experiences of a nonclinical adult sample across an eleven week period.  Each evening across an $11$-week period participants evaluated their daily emotional experiences using the modified Differential Emotions Scale (mDES) \citep{fredrickson2013}. The mDES is a $20$-item measure representing ten positive emotions and ten negative emotions. For the purpose of our current study we included all $10$ indicators for each of the emotion constructs. The question of how best to handle missing data within penalized estimation framework is an open question and currently missing data routines are not supported in the multi-VAR framework. For this reason we retained subjects with less than $10\%$ missing data and imputed the missing values component-wise using the predicted values from a single run of the Kalman Filter. This procedure left us with $16$ subjects on which to conduct our analysis.  

For each of these $16$ individuals in our sample we partitioned the data matrix into a training and test set.  The training set contained the first $77$ days of the $82$ day observation period and was used to estimate the various model parameters. The test set contained the final $5$ days of the observation period and was used to evaluate the accuracy of the different methods.  In addition to the individual-level LASSO and multi-VAR approaches (standard and adaptive)  we also considered some benchmark forecasting methods.  These methods include (a) the series \emph{average} where all future forecasts are equal to the mean of the training data, (b) a \emph{na\"{i}ve} method where all forecasts are set to the value of the last observation in the training set, (c) a \emph{drift} method which consists of drawing a straight line between the first and final observation of the training set, and extrapolating that trend line into the test set, (d) an \emph{AR(1)} model fit to each component of the training series and (e) an unrestricted \emph{VAR(1)} model. Root Means Squared Forecast Error was used to evaluate forecasts for each of the $5$ forecast horizons. Both the individual-level LASSO and multi-VAR approaches require tuning the $\lambda$ regularization parameters. To select the optimal $\lambda$ values we used the rolling-window cross-validation approach described previously.  

\begin{figure}[h]
\caption{Results from \citet{fredrickson2017} Data Across Approaches}
\label{fig:emp1}
\centering
\includegraphics[width=1\textwidth]{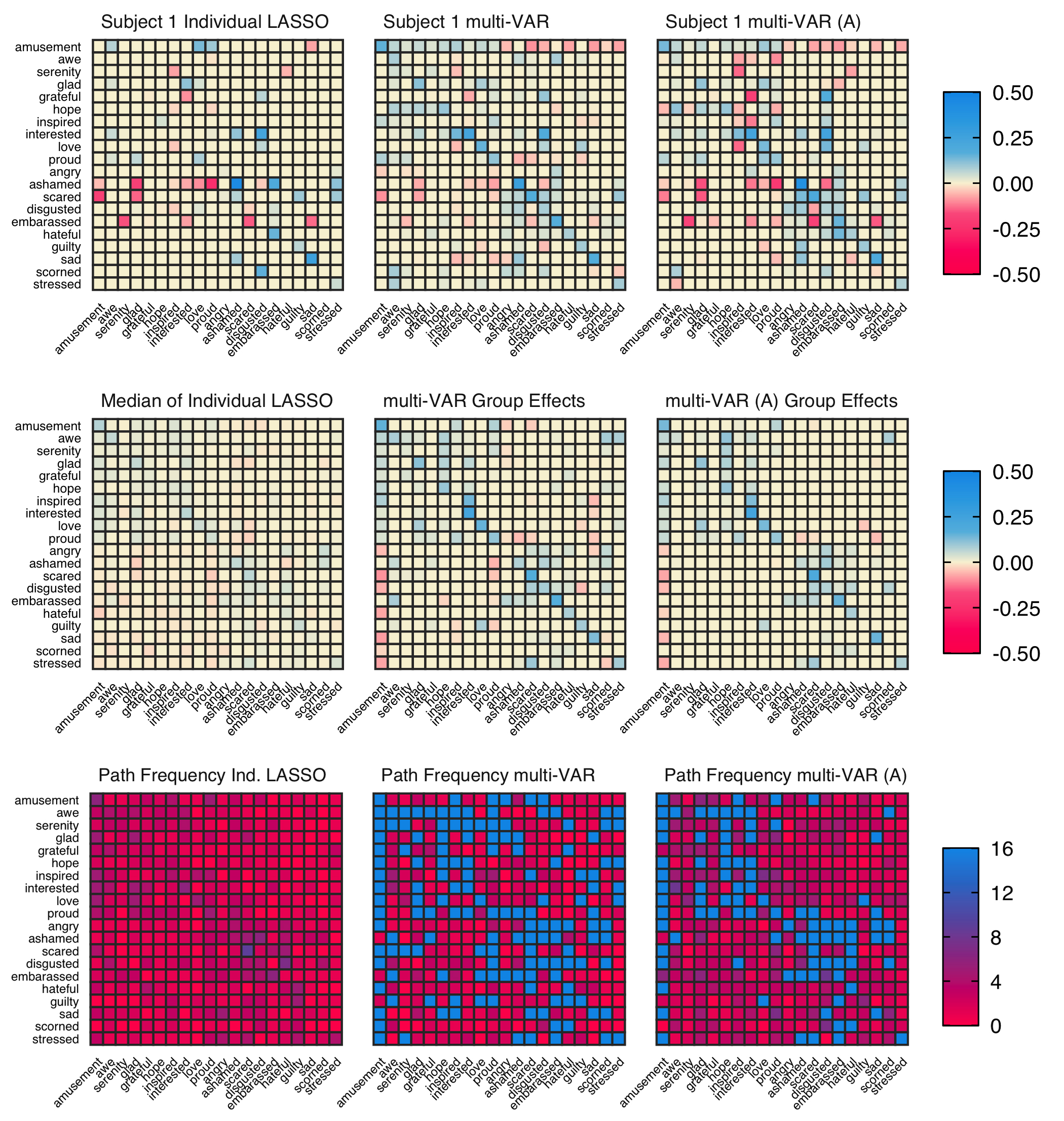}
\end{figure}

The $1-5$-step ahead forecast accuracy for the individual methods are given in Table \ref{emp1}. The LASSO-based approaches performed similarly and obtained the smallest forecast error across the forecasting approaches we evaluated. Within the LASSO approaches the two multi-VAR approaches performed the best in aggregate. In addition, Figure \ref{fig:emp1} provides a snapshot of the recovered transition matrices across the three approaches. In the first row of the figure the transition matrices for Subject 1 from each algorithm are shown.  The second row of Figure \ref{fig:emp1} provides a comparison of common effects resolved from the different approaches. For the individual LASSO the matrix represents the median effects across all individuals. For the multi-VAR approaches the transition matrices are the common effect matrices obtained from the algorithm directly. Lastly, for each method the third row provides the path frequency counts across all individuals in the sample. Here one can see a similar pattern of sparsity, as well as clustering within the positive and negative sub-scales. This is consistent with previous studies  which have used a bivariate dynamic factor analysis approach to model positive and negative items from the mDES as representing distinct but interdependent constructs \citep{fisher2020}.

\begin{table}[ht]
\centering
\caption{Root Mean Squared Forecast Error for \citet{fredrickson2017} Data} 
\label{emp1}
\scalebox{1}{
\begin{threeparttable}
\begin{tabular}{rccccc}
  \toprule
  &    \multicolumn{5}{c}{Forecast Window Length}  \\
\cmidrule(lr){2-6} 
 \multicolumn{1}{c}{Method} &
        \multicolumn{1}{c}{1} &
        \multicolumn{1}{c}{2} &
        \multicolumn{1}{c}{3} &
        \multicolumn{1}{c}{4} &
        \multicolumn{1}{c}{5} \\
 \midrule
Mean & 0.84 & 0.88 & 1.00 & 1.05 & 0.98 \\ 
Na\"{i}ve & 0.89 & 1.15 & 1.34 & 1.31 & 1.35 \\ 
Drift & 0.89 & 1.17 & 1.36 & 1.34 & 1.39 \\ 
AR(1) & 0.79 & 0.88 & 1.01 & 1.05 & 0.98 \\ 
VAR(1) & 0.83 & 0.90 & 1.03 & 1.04 & 0.97 \\ 
LASSO & 0.82 & 0.86 & 1.00 & 1.02 & 0.94 \\ 
multi-VAR & 0.75 & 0.85 & 0.97 & 1.03 & 0.95 \\ 
multi-VAR (A) & 0.76 & 0.85 & 0.97 & 1.03 & 0.95 \\ 
  \bottomrule
  \end{tabular}
   \begin{tablenotes}
   \item Note. (A) indicates adaptive multi-VAR.
 \end{tablenotes}
\end{threeparttable}
}
\end{table}

\section{Discussion}

This paper presents a novel approach for synthesizing multivariate time series obtained from multiple individuals. This method is especially well-suited to ILD paradigms when it is unclear how much individuals differ in terms of their dynamic processes. If individuals share little in common results from the proposed method resemble what would be obtained from fitting separate models to each individual. If individuals are homogenous results resemble what would be obtained from pooling the data and fitting a single model to the sample. Most importantly, if the truth lies somewhere in between these extremes - certain dynamics are shared while others are idiosyncratic - results will reflect this and provide researchers with new tools for isolating generalizable dynamics. Importantly, the simulation results presented here confirm that across three different levels of cross-sectional heterogeneity the proposed methods perform well in recovering the model dynamics and forecasting compared to benchmark methods.\\  

Despite these developments a number of limitations and opportunities for future development are worth considering. First, although we addressed some limitations of the VAR modeling in the context of ILD, others remain. For example, we assume the parameters themselves do not vary across time. This may be a strong assumption in the context of emotional dynamics. Second, in our simulation study and empirical example we set the regularization parameter $\lambda_{2}$ to be constant across individuals. In practice this is unlikely to hold and relaxing this assumption would potentially lead to better performance for the multi-VAR approaches, at an increased computational cost.\\

Relatedly, a limitation of our current work and an important area for future development involves identifying alternative methods for resolving the optimal multi-VAR penalty parameters. It was noted in the simulation study performance of the adaptive multi-VAR approach degraded considerably when the penalty parameters were chosen by RWCV.   Although there appears to be little consensus on which performance estimation method works best in the case of time-series data, two approaches are often considered: (1) Out-of-sample (OOS), and (2) cross-validation (CV) methods \citep{cerqueira2020}. Choosing an appropriate method depends on the specific characteristics of the data. The difference between OOS and CV methods is that OOS methods always preserve the temporal order of the observations and a model is never tested on historical data, relative to the training data. CV approaches, such as K-folds CV, often break the temporal order of time series and may produce poor estimates of predictive performance in real time-series contexts.

However, recent work has empirically demonstrated that CV methods perform well for stationary time series data, even outperforming OOS approaches in some circumstances \citep{bulteel2018a, bergmeir2012, bergmeir2014, bergmeir2018}. The reasons for this are not entirely clear although a number of adaptations to traditional K-fold CV have been made to accommodate time dependence through block-sampling (see \citet{bulteel2018a} for a review).  One explanation may be that CV approaches more efficiently use the available data, without requiring hold-out or initialization samples. CV approaches may be highly relevant for the smaller sample sizes considered here and the adaptive multi-VAR framework. Future work should explore which if any of the existing OOS or CV approaches are particularly well suited for the multi-VAR construction.

Another important area of development is to compare the multi-VAR approach to other frameworks capable of handling multiple-subject time series data. Two prominent methods are multilevel time series modeling  \citep{bringmann2013a, epskamp2018a} and GIMME \citep{gates2012}. While the current procedure relies on the VAR model, GIMME is based on the Structural VAR, making a naive comparison difficult. Work is currently under way to extend multi-VAR to the Structural and Graphical VAR frameworks. It is our hope these extensions will provide researchers with more tools for flexibly accommodating cross-sectional dependence in time-series data.

Based on the described results approaches capable of accommodating individual idiosyncrasies while exploiting what is common hold great promise for improving our ability to characterize and forecast complex physical and mental health outcomes at the individual level. In this vein we are optimistic the continued adoption of forecasting methodology by social and behavioral science researchers will only help to further integrate the nomothetic and idiographic approaches.  

\section{Appendix}
In this technical appendix, we discuss some theoretical aspect of LASSO estimation in the multi-VAR setting, namely, concerning its consistency and sparsistency.

\textbf{Consistency}: Consistency of LASSO estimation for single (stable) VAR models was established in the seminal paper by \cite{basu2015a,basu2015b}, building upon such results in the regression setting by \citet{loh2012a,loh2012b}. In the multi-VAR setting, the model is inherently unidentifiable. It could be that the LASSO solution is consistent for some particular $ \boldsymbol{\mu}^{*}$, $\boldsymbol{\Delta}^{*}_{k}$ in the model \eqref{decomp}, or over a subset of such identifications, but this problem still appears largely unresolved. Some related result though can be found in the discussion on sparsistency below following \cite{ollier2017}. Here, we shall discuss a weaker form of consistency of $\hat{\mathbf{B}}_{k}=\hat{\boldsymbol{\mu}}+\hat{\boldsymbol{\Delta}}_{k}$ for $\mathbf{B}_k^{*}$. The arguments are quite straightforward and shed some light on the problem, and also seemingly were not made in the related literature yet. 

We first describe the basic result for a single VAR model expressed in the regression form \eqref{eq5}, and then turn to a multi-VAR model. We index the model quantities with subscript $k$ or superscript $(k)$, $k=1,\ldots,K$, representing the individual models in the multi-VAR setting. After expanding the quadratic term of the objective function \eqref{lasso}, the estimation equation can be rewritten as in \citet{basu2015a} in terms of the quantities
\begin{equation}
\label{quantities}
\widehat{\Gamma}_{k}=\frac{1}{N}\mathbf{Z}^{(k)'}\mathbf{Z}^{(k)}=\frac{1}{N}(\mathrm{I}_{d}\otimes\mathcal{X}^{(k)}\mathcal{X}^{(k)'}),\quad \hat{\gamma}_{k}=\frac{1}{N}\mathbf{Z}^{(k)'}\mathbf{Y}^{(k)}.
\end{equation}
Estimation consistency is proved under the following two conditions on these quantities:

\begin{itemize}
\item Restricted eigenvalue condition: The matrix $\widehat{\Gamma}_{k}$ is said to satisfy this condition with parameters $\alpha_{k},\tau_{k}>0$,  if
\begin{equation}
\label{re_condition}
\beta_{k}'\widehat{\Gamma}_{k}\beta_{k}\geq\alpha_{k}\|\beta_{k}\|_{2}^{2}-\tau_{k}\|\beta_{k}\|_{1}^{2},\quad \beta_{k}\in\mathbb{R}^{q},
\end{equation}
with $q=pd^{2}$.

\item Deviation condition: This condition is satisfied if
\begin{equation}
\|\hat{\gamma}_{k}-\widehat{\Gamma}_{k}\mathbf{B}_{k}^{*}\|_{\infty}  \leq Q_{k}(\mathbf{B}_{k}^{*},\Sigma_{k,\varepsilon}) \sqrt{\frac{\log q}{N}},
\end{equation}
for a deterministic function $Q_{k}$.
\end{itemize}

Let $s_{k}=\|\mathbf{B}_{k}^{*}\|_{0}$ denote the sparsity of the model. Under the conditions above and assuming $s_{k}\tau_{k}\leq \alpha_{k}/32$, Proposition 4.1 of \citet{basu2015a} states that any solution $\hat{\mathbf{B}}$ of \eqref{lasso} satisfies: for any $\lambda \geq 4Q_{k}(\mathbf{B}_{k}^{*},\Sigma_{k,\varepsilon})\sqrt{\frac{\log q}{N}}$,
\begin{equation}
\label{consistency}
\|\hat{\mathbf{B}}_{k}-\mathbf{B}_{k}^{*}\|_{1}\leq \frac{64s_{k}\lambda}{\alpha_{k}},\quad \|\hat{\mathbf{B}}_{k}-\mathbf{B}_{k}^{*}\|_{2}\leq \frac{16\sqrt{s_{k}}\lambda}{\alpha_{k}}.
\end{equation}
Additionally, a result on the support of thresholded estimators of $\hat{\mathbf{B}}_{k}$ is also available. The consistency results in \eqref{consistency} apply to generic LASSO estimators as long as the quantities $\widehat{\Gamma}_{k},\hat{\gamma}_{k}$ satisfy the restricted eigenvalue and deviation conditions. 

Among the key contributions of \citet{basu2015a} are their results (Propositions 4.2 and 4.3) proving that $\widehat{\Gamma}_{k}$ and $\hat{\gamma}_{k}$ satisfy the restricted eigenvalue and deviation conditions with high enough probabilities, and expressing the various parameters involved in the conditions $(\alpha_{k},\tau_{k},Q_{k}(\mathbf{B}_{k}^{*},\Sigma_{k,\varepsilon}))$ in terms of the VAR model parameters. Furthermore, in the restricted eigenvalue condition, $\tau_{k}$ can be chosen so that $s_{k}\tau_{k}\leq \alpha_{k}/32$. We also note that the right-hand side of the inequalities \eqref{consistency} are expected to be negligible for small $\lambda$ and hence small $\log q/N$. The case when the logarithm of the dimension compares to the sample size through this way is the typical LASSO scenario.

In the multi-VAR setting, the optimization problem \eqref{lasso2} can be expressed through the objective function
\begin{equation}
\label{objectivefunction}
-\sum_{k=1}^{K}2\mathbf{B}_k'\hat{\gamma}_k + \sum_{k=1}^{K}\mathbf{B}_k'\widehat{\Gamma}_{k}\mathbf{B}_k + \lambda_{1}\|\boldsymbol{\mu}\|_1 +  \sum_{k=1}^{K}\lambda_{2,k}\|\mathbf{B}_{k}-\boldsymbol{\mu}\|_1.
\end{equation}
A consistency bound for the minimizer $\hat{\mathbf{B}}_{k}$ of \eqref{objectivefunction} can still be obtained similarly as for single VAR models if one is willing to make the assumption
\begin{equation}
\label{assump_consist}
\|\hat{\boldsymbol{\mu}}\|_0 \leq s_0.
\end{equation}
The constraint \eqref{assump_consist} could be imposed while optimizing \eqref{objectivefunction} or choosing $\lambda_1$ appropriately large, or inferred to hold (with high enough probability) from sparsistency result, if available. Indeed, under \eqref{assump_consist}, a consistency bound can be derived easily as in the proof of Proposition 3.3 in \cite{basu2015a,basu2015b}. That is, observe first that
\begin{eqnarray*}
&& -\sum_{k=1}^{K}2\hat{\mathbf{B}}_k'\hat{\gamma}_k + \sum_{k=1}^{K}\hat{\mathbf{B}}_k'\widehat{\Gamma}_{k}\hat{\mathbf{B}}_k + \lambda_{1}\|\hat{\boldsymbol{\mu}}\|_1 +  \sum_{k=1}^{K}\lambda_{2,k}\|\hat{\mathbf{B}}_{k}-\hat{\boldsymbol{\mu}}\|_1 \\
&\leq& -\sum_{k=1}^{K}2\mathbf{B}_k^{*'}\hat{\gamma}_k + \sum_{k=1}^{K}\mathbf{B}_k^{*'}\widehat{\Gamma}_{k}\mathbf{B}_k^{*} + \lambda_{1}\|\hat{\boldsymbol{\mu}}\|_1 +  \sum_{k=1}^{K}\lambda_{2,k}\|\mathbf{B}_{k}^{*}-\hat{\boldsymbol{\mu}}\|_1
\end{eqnarray*}
and rearranging the terms and setting $\mathbf{v}_k=\hat{\mathbf{B}}_k-\mathbf{B}_k^*$, we deduce
\begin{equation*}
\sum_{k=1}^{K}\mathbf{v}_k \widehat{\Gamma}_k \mathbf{v}_k \leq \sum_{k=1}^{K}2\mathbf{v}_k' (\hat{\gamma}_k-\widehat{\Gamma}_k \mathbf{B}_k^{*})+\sum_{k=1}^{K}\lambda_{2,k}\left( \|\mathbf{B}_k^{*}-\hat{\boldsymbol{\mu}}\|_1 - \|\mathbf{B}_{k}^{*} - \hat{\boldsymbol{\mu}} + \mathbf{v}_k\|_1 \right).
\end{equation*}
With $\hat{J}_{k}=\textrm{supp}\{ \mathbf{B}_k^{*} - \hat{\boldsymbol{\mu}}$ being the index support of $\mathbf{B}_k^{*} - \hat{\boldsymbol{\mu}}\}$, repeating the argument in \cite{basu2015a,basu2015b}, we get
\begin{equation}
\label{result_from_basu}
0 \leq \sum_{k=1}^{K} \mathbf{v}_k' \widehat{\Gamma}_k \mathbf{v}_k \leq \sum_{k=1}^{K}\left( \frac{3\lambda_{2,k}}{2}\|(\mathbf{v}_{k})_{\hat{J}_k}\|_1 -\frac{\lambda_{2,k}}{2}\|(\mathbf{v}_{k})_{\hat{J}_k^{c}}\|_1  \right)
\end{equation}
as long as $\lambda_{2,k}\geq4Q_k(\mathbf{B}_k^{*},\Sigma_{k,\varepsilon})\sqrt{\frac{\log q}{N}}$ (with the function $Q_k$ from the deviation condition), where $(\cdot)_{\hat{J}}$ and $(\cdot)_{\hat{J}^c}$ denote restrictions to the index sets $\hat{J}$ and $\hat{J}^c$, respectively. Then,
\begin{equation*}
\sum_{k=1}^{K}\lambda_{2,k}\|(\mathbf{v}_k)_{\hat{J}_k^{c}}\|_1 \leq 3\sum_{k=1}^{K}\lambda_{2,k}\|(\mathbf{v}_k)_{\hat{J}_k}\|_1
\end{equation*}
and one also has
\begin{eqnarray}
&& \sum_{k=1}^{K}\lambda_{2,k}\|\mathbf{v}_k\|_1 \leq 4\sum_{k=1}^{K}\lambda_{2,k}\|(\mathbf{v}_k)_{\hat{J}_k}\|_1 \nonumber\\
&\leq& 4\sum_{k=1}^{K}\lambda_{2,k}(s_0+s_k)^{1/2}\|\mathbf{v}_k\|_2 \leq 4 \sqrt{\sum_{k=1}^{K}\lambda_{2,k}^{2}(s_0+s_k)}\|\mathbf{v}\|_2,
\end{eqnarray}
by Cauchy-Schwarz inequality (twice) and the fact that $|\textrm{supp}\{\hat{J}_{K}\}|\leq s_0+s_k$, where $\|v\|_2^2=\sum_{k=1}^{K}\|v_k\|_2^2$. Similarly, by the restricted eigenvalue condition \eqref{re_condition} for each model and assuming $s_k\tau_k \leq \alpha_k/32$, we have
\begin{equation}
\label{result_from_re}
\sum_{k=1}^{K} \mathbf{v}_k'\widehat{\Gamma}_k \mathbf{v}_k \geq \sum_{k=1}^{K}\frac{\alpha_k}{2}\|\mathbf{v}_k\|_2^2 \geq \frac{\min\{\alpha_k\}}{2}\|\mathbf{v}\|_2^2.
\end{equation}
A combination of \eqref{result_from_basu}-\eqref{result_from_re} yields e.g.
\begin{equation}
\frac{\min\{\alpha_k\}}{2}\|\mathbf{v}\|_2^2 \leq 6\sqrt{\sum_{k=1}^{K}\lambda_{2,k}^2(s_0+s_k)}\|\mathbf{v}\|_2
\end{equation}
or
\begin{equation}
\|\mathbf{v}\|_2 \leq \frac{12\sqrt{\sum_{k=1}^{K}\lambda_{2,k}^2(s_0+s_k)}}{\min\{\alpha_k\}}.
\end{equation}
This is the multi-VAR analogue of the second consistency bound in \eqref{consistency}. One can similarly obtain a bound on $\|\mathbf{v}\|_1$ analogous to the first one in \eqref{consistency}.\\
\textbf{Sparsistency}: We comment here briefly on the possibility of recovering the supports of $\boldsymbol{\mu}^{*}$ and $\boldsymbol{\Delta}_{k}^{*}$. The same issue of (non)identifiability is fundamental here as well. Some result nevertheless are available in the literature for special cases. Assuming effectively that $s\lambda_1/\lambda_{2,k}=cK^{1/2}$, \cite{ollier2017} gave conditions for identifiability and sparsistency with the limiting common parameter of interest $\boldsymbol{\mu}^{*}$ defined as the entrywise median of $\boldsymbol{B}_k^{*}$. Their approach goes through verifying a particular well-known irrepresentability condition on a design matrix. It could in principle be adapted to the multi-VAR context but the value of this effort might be questionable. First, irrepresentability conditions are quite restrictive and difficult to verify, and as a result, adaptive LASSO versions are advocated for. The setting where the limiting parameter of interest is necessarily related to the median could also be viewed restrictive.

\bibliography{multivar_paper}
\end{document}